\documentclass[smus]{snow2e}
\usepackage[dvips]{graphics}

\def\beq{\begin{equation}}
\def\eeq#1{\label{#1}\end{equation}}
\def\eeqn{\end{equation}}
\def\beqa{\begin{eqnarray}}
\def\eeqa#1{\label{#1}\end{eqnarray}}
\def\eeqan{\end{eqnarray}}
\def\CR{\nonumber \\ }
\def\leqn#1{(\ref{#1})}
\def\bar#1{\overline{#1}}
\def\Dslash{\not{\hbox{\kern-4pt $D$}}}
\def\half{{1\over 2}}
\def\L{{\cal L}}

\def\tr{{\rm tr}}
\def\del{\partial}

\def\VEV#1{\left\langle{ #1} \right\rangle}
\def\ee{e^+e^-}
\def\sstw{\sin^2\theta_w}
\def\cstw{\cos^2\theta_w}
\def\etal{{\it et~al.}}

\newcommand\pubnumber{SLAC-PUB-7397\\ CLNS 97/1473}
\newcommand\pubdate{April, 1997}

\def\Title#1{\begin{center} {\Large #1 } \end{center}}
\def\Author#1{\begin{center}{ \sc #1} \end{center}}

\def\doeack{\footnote{Work supported by the Department of Energy,
                     contract DE--AC03--76SF00515.}}
\def\burdack{\footnote{Work supported in part by the Department of Energy,
                     contract DE--FG02--95ER40896.}}
\def\hadack{\footnote{Work supported in part by the Department of Energy,
                     contract DE--FG05--91ER40670.}}
\def\buack{\footnote{Work supported in part by the Department of Energy,
                     contract DE--FG02--91ER40676.}}
\def\drellack{\footnote{Work supported in part by the National Science
              Foundation,  contract PHY-93-10754.}}
\def\ternack{\footnote{Work supported in part by the National Science
              Foundation,  contract PHY-95-14797.}}
\def\woodack{\footnote{Work supported in part by the National Science
              Foundation,  contract PHY-95-14950.}}
\def\fermiack{\footnote{Work supported by the Department of Energy,
                     contract DE-AC02-76HO3000.}}

\newcommand\pubblock{\rightline{\begin{tabular}{l} \pubnumber\\
         \pubdate  \end{tabular}}}
\newenvironment{Abstract}{\begin{quotation} \begin{center}
                       ABSTRACT
     \end{center}\bigskip  }{\end{quotation}}
\begin{document}

\onecolumn

\hoffset +1mm
\voffset +5mm

\begin{titlepage}
\pubblock

\vfill

\Title{Strong Coupling Electroweak Symmetry 
Breaking}

\vfill
\Author{Timothy L. Barklow\doeack\\
{\it Stanford Linear Accelerator Center, Stanford University, Stanford CA
94309}\\[.6ex]
Gustavo Burdman\burdack\\{\it Department of Physics, University of 
 Wisconsin, Madison, WI 53706}\\[.6ex]
R. Sekhar Chivukula and Bogdan A. Dobrescu\buack\\
 {\it Department of Physics, Boston University, Boston, MA
 02215}\\[.6ex]
Persis S. Drell\drellack\\
{\it Laboratory of Nuclear Studies, Cornell
 University, Ithaca, NY 14853-5001}\\[.6ex] 
Nicholas Hadley\hadack\\{\it Department of Physics, University of 
Maryland, College Park, MD 20742}\\[.6ex] 
William B. Kilgore\fermiack\\{\it Fermi National Accelerator Laboratory,
           Batavia, IL 60510}\\[.6ex]
Michael E. Peskin\doeack\\{\it  Stanford Linear 
Accelerator Center, Stanford University, Stanford CA
94309}\\[.6ex]
John Terning\ternack\\{\it Department of Physics, University of California, 
       Berkeley, CA 94720}\\[.6ex]
Darien R. Wood\woodack\\{\it Department of Physics, Northeastern University,
          Boston, MA 02115}}
\vfill
\begin{center}
Working Group Summary Report from the 1996 DPF/DPB Summer Study\\
New Directions for High Energy Physics\\
Snowmass, Colorado, June 25--July 12, 1996
\end{center}
\vfill
\end{titlepage}
\newpage
\begin{titlepage}
\mbox{\null}

\vfill
\begin{Abstract}
We review models of electroweak symmetry breaking due to new strong
interactions at the TeV energy scale and discuss the prospects for
their experimental tests.  We emphasize the direct observation of the
new interactions through high-energy scattering of vector bosons.  We
also discuss indirect probes of the new interactions and exotic
particles predicted by specific theoretical models. 
\end{Abstract}

\vfill
 \setcounter{footnote}{0}
\end{titlepage}

\setlength{\textheight}{241mm}   
\setlength{\textwidth}{182mm}
\setlength{\columnsep}{4mm}
\setlength{\topmargin}{7mm} 
\setlength{\oddsidemargin}{20mm}
\setlength{\evensidemargin}{20mm}

\hoffset -28mm
\voffset -32mm

\twocolumn

\title{Strong Coupling Electroweak Symmetry 
Breaking\thanks{This work was supported in part by the U. S. Department of 
Energy and by the National Science Foundation.}}

\author{Timothy L. Barklow\\
{\it Stanford Linear Accelerator Center, Stanford University, Stanford CA
94309}\\[.6ex]
Gustavo Burdman\\{\it Department of Physics, University of 
 Wisconsin, Madison, WI 53706}\\[.6ex]
R. Sekhar Chivukula and Bogdan A. Dobrescu\\
 {\it Department of Physics, Boston University, Boston, MA
 02215}\\[.6ex]
Persis S. Drell\\
{\it Laboratory of Nuclear Studies, Cornell
 University, Ithaca, NY 14853-5001}\\[.6ex] 
Nicholas Hadley\\{\it Department of Physics, University of 
Maryland, College Park, MD 20742}\\[.6ex] 
William B. Kilgore\\{\it Fermi National Accelerator Laboratory,
           Batavia, IL 60510}\\[.6ex]
Michael E. Peskin\\{\it  Stanford Linear 
Accelerator Center, Stanford University, Stanford CA
94309}\\[.6ex]
John Terning\\{\it Department of Physics, University of California, 
       Berkeley, CA 94720}\\[.6ex]
Darien R. Wood\\{\it Department of Physics, Northeastern University,
          Boston, MA 02115}}

\maketitle

\begin{abstract} 
We review models of electroweak symmetry breaking due to new strong
interactions at the TeV energy scale and discuss the prospects for
their experimental tests.  We emphasize the direct observation of the
new interactions through high-energy scattering of vector bosons.  We
also discuss indirect probes of the new interactions and exotic
particles predicted by specific theoretical models. 
\end{abstract}

\section{Introduction}
Though it is often said that the experiments of the last six years at
LEP, SLC, and the Tevatron have brought no surprises, this very fact
has led us into a new era in our understanding of particle physics.  In
the past, it has been possible to regard the $SU(2)\times U(1)$ gauge
theory of the weak and electromagnetic interactions as a provisional
theory, perhaps to be replaced by a model in which $W$ bosons have
constituents or internal structure. But the new experiments on the
detailed properties of the $Z$ and $W$ bosons have confirmed the gauge
theory predictions at the level of loop corrections. A striking aspect
of this confirmation is the agreement between the value of the top
quark needed to give the proper radiative corrections and the value of
the top quark mass actually observed by the Tevatron collider
experiments. The data compel us to accept $SU(2)\times U(1)$ as a
fundamental gauge symmetry of Nature, a symmetry on the same footing as
the gauge symmetry of electromagnetism.

At the same time, our increased understanding of the theory of
electroweak interactions highlights the one central unsolved problem of
that theory. In order that the $W$ and $Z$ bosons acquire mass, the
$SU(2)\times U(1)$ gauge symmetry must be spontaneously broken.  What
{\em causes} this spontaneous symmetry breaking? At the moment, we have
almost no experimental clue that bears on this question.  In principle,
the symmetry breaking may be caused by an elementary scalar field (the
Higgs field) obtaining a vacuum expectation value, or by the vacuum
expectation value of a composite operator.  At this moment, it is more
fashionable to assume that the symmetry breaking is caused by an
elementary scalar field with only weak-coupling interactions.  This
viewpoint connects naturally to supersymmetry, which Marciano, in his
introductory lecture at this meeting \cite{Marcianointro}, called `the
only good idea out there'.  On the other hand, it is attractive
intuitively that an important rearrangement of symmetry such as we know
occurs at the electroweak scale should result from new strong
interactions.  In this article, we will take this as our fixed idea and
review its consequences in detail.

What, then, are the consequences of new strong interactions responsible
for electroweak symmetry breaking?  How will we investigate these new
interactions experimentally? In particular, how much will we learn
about them at the next generation of colliders?

Our discussion of these issues will proceed as follows:  In Section II,
we will review the present, rather weak, constraints on the nature of
these new interactions.  Sections III--VI, the heart of this review,
will discuss direct experimental probes of the new sector.  These
necessarily are experiments at very high energy, requiring also very
high luminosity. In this sense, they test the ultimate reach of the
colliders we are planning for the next generation, and for the more
distant future. In Section III, we will present a general
phenomenological theory of  new strong interactions at the TeV scale,
and we will raise a definite set of questions for experiment to
address. In Section IV, we will review proposed experiments on the
scattering of weak vector bosons through the new strong interactions. 
In Section V, we will discuss an experimental probe of the top quark's
connection to the new interactions. In Section VI, we will summarize
the sensitivity of these experiments to the resonances of the new
strong interactions, and the complementarity of different experimental
probes.

In Sections VII-IX, we will discuss additional consequences of the new
strong interactions which are potentially accessible to experiments at
lower energies.  In Section VII, we will discuss possible anomalous
gauge boson couplings and the manner in which these probe new strong
interactions. In Section VIII, we review some more explicit models of
strong-coupling electroweak symmetry breaking and explain how their
specific dynamical assumptions lead to the prediction of exotic
particles and interactions.  In Section IX, we review experiments at
present and future colliders which can search for the new particles
predicted in these models.

Finally, in Section X, we summarize our discussion and present our
general conclusions.

The reader experienced in the study of strong-coupling electroweak
symmetry breaking will find relatively little new material in this
review. Nevertheless, we feel that our time at Snowmass has been well
spent. The members of our working group arrived at Snowmass with
different types of expertise and very different preconceptions. We have
appreciated the opportunity to discuss our areas of disagreement and to
note how our  ideas can fit together into a  coherent overall picture.
We hope that our improved understanding of this subject will be
reflected here both in a clearer presentation of the main aims and
directions of the study of new strong interactions and in a clearer
understanding of the strengths and weaknesses of the experimental tools
we hope eventually to bring to this problem.

\section{Experimental Constraints on New Strong Interactions}

In this section, we discuss the present experimental constraints on 
new strong interactions responsible for electroweak symmetry breaking,
and an important additional constraint that will come from future
experiments.

\subsection{Present Constraints}

We begin with the information on the new strong sector that we have
today. This will be a very short section.  On the other hand, it is
important to realize that some definite constraints are available. In
particular, we know:

\begin{quote}
1.)  The new strong interaction sector has an $SU(2)\times U(1)$ global
symmetry, which it breaks spontaneously to $U(1)$. 
\end{quote}
That much is required
just to couple it to the standard model gauge fields.
\begin{quote}
2.)  The new strong interaction sector has an $SU(2)$ global symmetry
which is not spontaneously broken.
\end{quote}
The argument for this is that the relation 
\beq
m_W/m_Z = \cos\theta_w
\eeq{rhorel} 
is satisfied to high accuracy; the violation of this relation, at a
level below 1\%\ depending on the definition of $\sstw$, is fully
accounted for by the standard electroweak radiative corrections.  At
the same time, the photon mass is zero.  To insure these two relations,
the gauge boson mass matrix, in the $SU(2)\times U(1)$ basis
$(A^1,A^2,A^3,B)$, must have the form
\beq
 m^2 = {v^2\over 4} \pmatrix{g^2  &  &  &  \cr 
                          & g^2 & & \cr
                           &  & g^2 & -gg' \cr
                         & & -gg' & (g')^2 \cr}  \ .
\eeq{WZmasses}
Notice that the block of this matrix acting on the three $SU(2)$ gauge
bosons is completely symmetric.  If this mass matrix originates from a
strong-coupling theory, this symmetry must reflect a property of that
theory.  The simplest possibility is an {\it unbroken} $SU(2)$ gauge
symmetry under which the three $SU(2)$ currents transform as an isospin
triplet \cite{Marvin,SSVZ}.  This symmetry is known as `custodial
$SU(2)$' symmetry.
\begin{quote}
3.)  The mass scale of the new strong interaction sector 
is given by  $v= 246$ GeV.
\end{quote}
The value of $v$ follows from the known values of the $W$ and $Z$
masses and the electroweak coupling constants, which are related by
\leqn{WZmasses}. The relation between $v$ and the positions of
strong-interaction resonances needs some extra discussion, which we
will supply below.
\begin{quote}
4.)  The new strong interactions are not just a scaled-up version of QCD.
\end{quote}
This conclusion follows from a more detailed examination of the
precision electroweak data.  These measurements  are often analyzed for
the effects of new physics by introducing parameters $S$, $T$, $U$
which represent the effects of new particles on the $Z$ and $W$ vacuum
polarization diagrams \cite{PandT}. In particular, $S$, a finite part
of the $Z$ field strength renormalization, can be predicted in quite a
clean way if one assumes that the new interactions resemble scaled-up
QCD. The result is $S = 0.3 \pm 0.1$.  This should be compared to the
value for $S$ which is obtained by fitting the deviation of the
electroweak data from  the standard model predictions for a Higgs boson
mass of 1 TeV \cite{LandEnew}:
\beq
S = - 0.26 \pm 0.16 \ .
\eeq{Svalue} 
The results are inconsistent at the 3 $\sigma$ level. We will discuss
the properties of models which ameliorate this problem in Section VII.

\subsection{No Light Higgs?}

There is a fifth piece of information which is not yet available but
which will have an important influence on this experimental program
when it does become known.  This is the question of whether there is a
light Higgs particle. We will emphasize in this report that the
experiments which will explore the structure of the new strong
interactions are difficult ones, requiring high luminosity and high
energy.  Thus it will be important to know in advance whether one
really needs this strong-interaction hypothesis to explain electroweak
symmetry breaking.  That question can be addressed experimentally.

The alternative to electroweak symmetry breaking through new strong
interactions is the possibility of electroweak symmetry breaking due to
the vacuum expectation values of one or more  light scalar fields. We
can rule out this alternative if these scalar fields are not discovered
in collider experiments. These issues are discussed in more detail in
the working group report on weakly coupled Higgs bosons \cite{Haber}.

First of all, we should ask whether there exist light scalar particles,
detectable, for example, through their $b\bar b$ or $\gamma\gamma$
decay modes.  The experiments which search for these particles have
been given a prominent role in discussions of all future colliders.  It
is likely that a light scalar, if it exists, will be discovered at the
LHC if not before. However, the simple existence of these particles is
not sufficient information, because there are several  ways to obtain
particles of this sort, as  bound states of more elementary
constituents, in models in which $SU(2)\times U(1)$ breaking arises
from new strong interactions.  We will discuss several examples in
Section VIII. Because of this, it is also important to determine
whether a light scalar particle which has been observed plays a direct
role in electroweak symmetry breaking.

This can be done by observing the coupling of the scalar to $ZZ$ or
$WW$. If we write an $SU(2)\times U(1)$ invariant effective Lagrangian,
with several neutral scalar particles $\phi_i$ assigned (for
simplicity) a common weak isospin $I^3$, the coupling of these scalars
to $WW$ and $ZZ$ is given by
\beq
   \Delta \L  = (2I^3)^2  \sum_i{w_i\over v^2} \left(m_Z^2 Z_\mu Z^\mu 
         + 2 m_W^2 W_\mu^+ W^{-\mu}
            \right) \phi_i  \ ,  
\eeq{Higgscouple}
where 
\beq
        w_i =  \sqrt{2}\VEV{\phi_i} \ .
\eeq{wval}
and $v$ is the scale given  in item 3 above.  If
\beq
       \sum_i  (2I^3)^2 w_i^2 = v^2 \ ,
\eeq{wsumrule}
then the vacuum expectation values of the fields $\phi_i$ are
completely responsible for the generation of the $Z$ and $W$ masses
through the breaking of $SU(2)\times U(1)$.  If particles are not found
which saturate the sum rule \leqn{wsumrule}, then the mechanism of
$SU(2)\times U(1)$ breaking necessarily acts at higher energies.
Because a Higgs particle can be heavy only if it is strongly self-coupled,
this mechanism would also necessarily involve new strong interactions.

It is difficult to measure the vacuum expectation values $w_i$ at
hadron colliders.  This coupling is best observed in the process $q\bar
q \to W \phi$, with $\phi$ decay to $b\bar b$.  For standard model
coupling ($w=1$), this process can be observed at the upgraded Tevatron
up to a $\phi$ mass of about 110 GeV, depending on the final integrated
luminosity \cite{TeV2000}, and over roughly the same range at the LHC
\cite{IanandJohn}.  At electron colliders, where the dominant process
for Higgs boson production is $\ee\to Z^0 \phi$, this experiment is
quite straightforward.  In particular, once the mass of the $\phi$ is
given, the measurement of $w$ involves only counting $Z^0$ bosons
produced at a fixed lab energy and can be done without assumptions on
the $\phi$ decay scheme \cite{NLCrep}. If the bosons are found which
saturate the sum rule \leqn{wsumrule}, these particles hold the physics
of electroweak symmetry breaking.  If they are not found, we will know
that we need the experimental program for very high energies that we
will set out below.

We should note that it is possible that a light Higgs boson responsible
for electroweak symmetry breaking could also be a composite bound by
new strong interactions.  Models of this type were introduced many
years ago by Kaplan and Georgi \cite{KandG} and have also appeared
recently in the context of supersymmetry model-building \cite{NS}.  In
this case, the scale of new strong interactions can lie at a multi-TeV
mass scale. We will not discuss these models further in this report.

\section{New Pion Dynamics}

Although we do not know much about the nature of the new strong
interactions, we know enough to suggest a general route for
experimental analysis.  In this section, we will set out a general
phenomenology of the new strong interactions which can provide a guide
for us in the discussion of experiments.

From items 
1 and 2 above, we know that the new strong interactions contain the 
pattern of spontaneous global symmetry breaking:
\beq 
    SU(2) \times SU(2) \to SU(2) \ ,
\eeq{twotwotwo}
which is identical to the symmetry breaking pattern of QCD in the limit
of zero mass for the $u$ and $d$ quarks.  From here on, to promote this
analogy, we will refer to this zero quark mass limit of QCD simply as
`QCD', and we will refer to the unbroken $SU(2)$ symmetry of the new
strong interactions as `isospin'. Then, for both the familiar and the
new theory of strong interactions,  Goldstone's theorem implies that
there should exist an isospin triplet of massless mesons, the pions
$(\pi^+, \pi^-,\pi^0)$. Further, the fact that these particles are
Goldstone bosons resulting from spontaneous symmetry breaking leads to
a systematic set of predictions for the low-energy behavior of their
interactions, predictions which are in fact valid experimentally after
a correction is made  for the effects of the nonzero quark masses.

In the familiar strong interactions, the formalism which gives these
predictions is known as `current algebra'.  Where current algebra
predicts a property of the familiar pions, the same property must hold
for the pions of the new strong interactions. However, outside the
domain controlled by current algebra,  the properties of these new
pions could be completely different from the QCD expectation.  How the
new pions behave at high energy, we must find out experimentally.

The idea that the new strong interactions, just like QCD, have pions as
their lightest states gives focus to the experimental program to
investigate these new interactions. It suggests that we approach the
study of these new interactions by asking the same questions that we
asked in the 1950's about low-energy QCD.  Of course, we should be
prepared that the {\em answers} to these questions may be completely
different.

\subsection{Who are the Pions?}

First of all, we must clarify where the new pions can  be found. How do
we create a beam of the new pions, or look for them in the final state
of a high-energy reaction?

In the new strong interactions, the spontaneously broken $SU(2)$ is
coupled to the gauge bosons of the standard model.  When the symmetry
breaks, the weak gauge bosons obtain mass.  Since a massless gauge
boson has two transverse polarization states, but a massive spin-1
boson has three polarization states, each of these bosons must acquire
an additional degree of freedom. It is well known that they do this by
absorbing the corresponding Goldstone boson resulting from the global
symmetry breaking. This is the essence of the Higgs mechanism.

\begin{figure}[htb]
\leavevmode
\begin{center}
\resizebox{2.5in}{!}{%
\includegraphics{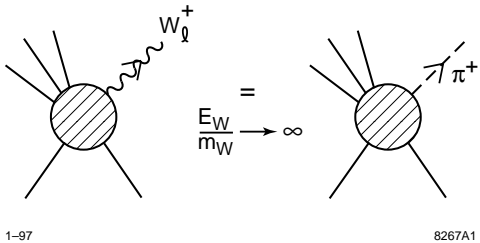}}
\end{center}
\caption{The Goldstone Boson Equivalence Theorem.}
\label{fig:gbet}
\end{figure}

This statement of the Higgs mechanism has an interesting converse. 
When a massive spin-1 boson is boosted to high energy, it again has
well-defined transverse and longitudinal polarization states.  The
transverse polarization states have couplings which approach those of
the original massless bosons. The interactions of the longitudinal
polarization states of the spin-1 boson become equal to those of the
Goldstone boson that it ate to become massive.  This result is known as
the `Goldstone boson equivalence theorem' \cite{GBET1,GBET3,CandG}; see
Figure~\ref{fig:gbet}.

The Goldstone boson equivalence theorem implies that the longitudinal
polarization states of $W$ and $Z$ play the role of the pions in the
new strong interactions.  Any collider process which involves $W$ and
$Z$ bosons in the initial or final state can in principle give us
access to these new interactions. In the next several sections, we will
discuss the most important reactions for the future study of these new
pion processes.

To complete the connection between $W$ and $Z$ dynamics and the
dynamics of new pions, we can add one more piece of information.  It is
possible to derive the $W$ and $Z$ mass matrix predicted by the new
strong interaction theory by analyzing  matrix elements of the
$SU(2)\times U(1)$ gauge currents. Assuming the symmetry-breaking
pattern \leqn{twotwotwo}, we must obtain a mass matrix of the form
\leqn{WZmasses}.  But this derivation also identifies the overall scale
$v$ in \leqn{WZmasses} as being equal to the pion decay constant:
\beq
            f_\pi = v = 246 \ \hbox{\rm GeV} \ .
\eeq{fpi}
Through this relation, we can convert the known masses of the $W$ and
$Z$ to the mass scale of the new strong interactions.

\subsection{Pion-Pion Scattering}

The most basic question that we could ask about the new strong
interactions is:  What is the nature and strength of the low-lying
resonances?  The analogy to pion dynamics in QCD tells us that we
should think especially about resonances that couple to two-pion
scattering channels.  From the scale given by \leqn{fpi}, we should
expect these `low-lying' resonances to be found at TeV energies.

In pion-pion scattering, Bose statistics dictates the principal
scattering channels at low energy.  These have total spin $J$ and
isospin $I$ given by:
\beqa
            J=0 &\qquad& I= 0, 2 \CR
            J=1 &\qquad&  I=1  \ .
\eeqa{IJs}
From now on, we will refer to these three channels by the value of $I$.
We will discuss in a moment how to look for resonances in these
channels using longitudinal $W$ and $Z$ bosons as our tools.

Current algebra predicts the leading behavior of the pion-pion
scattering amplitude near threshold.  Specifically, it predicts
\beq
             a_I = {s\over A_I} + {\cal O}(s^2)
\eeq{aasym}
where $A_I$, the relativistic generalization of  the scattering length,
is given in the three relevant channels by
\beqa
A_0 &=& 16\pi f_\pi^2 = (1.7\ \hbox{\rm TeV})^2 \CR
A_1 &=& 96\pi f_\pi^2 = (4.3\ \hbox{\rm TeV})^2 \CR
A_2 &=& -32\pi f_\pi^2  \ .
\eeqa{Avalues}

The channels $I$=0 and 1 are attractive at low energy, and so one might
expect to see resonances here.  In fact, a more powerful statement is
possible. Unitarity implies that Im($a_I) = |a_I|^2$, which implies
that Re($a_I) < 1/2$.  The corrections to \leqn{aasym} must become
important and restore $a_I(s)$ to an expression consistent with
unitarity before this criterion is met \cite{GBET3}, that is, for
\beq
  I = 0:\ \sqrt{s}< 1.3\ \hbox{\rm TeV} \ , \quad
 I = 1:\ \sqrt{s}  < 3.0\ \hbox{\rm TeV} \ .
\eeq{ulimits}
If unitarity is restored through resonances, as it is in the familiar
strong interaction, we would expect the first resonance in each channel
to appear below these limits. For comparison, the most prominent
low-energy resonance in QCD is the $\rho$.  Scaling up $m_\rho$ by the
ratio of $f_\pi$'s, we find
\beq
           m_\rho^{\hbox{\rm new}} =
 {f_\pi^{\hbox{\rm new}}\over f_\pi^{\hbox{\rm old}}}  m_\rho^{\hbox{\rm old}}
 = {246 \ \hbox{\rm GeV}
     \over 93 \ \hbox{\rm MeV}}  m_\rho = 2 \ \hbox{\rm TeV}
\eeq{rhomass}
as an alternative estimate of the $I$=1 resonance mass. 

In the literature, a prominent $I$=$J$=1 resonance in the new strong
interactions is called a `technirho'.  By analogy, a prominent $I$=0,
$J$=1 resonance is called a `techniomega'.  As in QCD, the techniomega
will not couple to $\pi\pi$, but it may appear in $\ee$ and  $q \bar q$
annihilation, decaying to $3 \pi$ or $\gamma \pi$ final states. 
Through the latter channel, the techniomega can appear as a resonance
in $\gamma Z^0$.  While the technirho, if it exists, must couple to a
virtual $\gamma$ or $Z^0$, for the techniomega this is a
model-dependent question.

When we compute the sensitivity of collider experiments to resonances
of the new strong interactions, and we consider parton-level processes
which extend above 1 TeV in center-of-mass energy, we must be careful
to respect unitarity. A direct, if simple-minded, way to do this is to
parametrize the {\em phase shift} in each of the three relevant
channels by the expression
\beq
         \tan\delta_I =  {s\over A_I} {1\over 1 - s/M_I^2} \ .
\eeq{standarddelta}
This expression automatically respects unitarity and the predictions of
current algebra.  It includes a resonance at the mass $M_I$.  It has a
sensible limit $M_I \to \infty$ which is determined only by current
algebra and unitarity.  In the literature, the limit of a unitarized
but nonresonant pion-pion scattering amplitude is known as the `low
energy theorem' (LET) model.  The precise definition of the LET limit
depends on the unitarization procedure and thus varies somewhat from
author to author.  All of these prescriptions give similar results for
$\pi\pi$ center-of-mass energies in the region \leqn{ulimits}.

The  simple parametrization \leqn{standarddelta} does not take account
of crossing relations which connect the amplitudes in the three
channels (and additional channels which become important at higher
energies).  The detailed calculations of the signatures of pion-pion
scattering processes in high-energy colliders \cite{Chano,bagger} have
used somewhat more sophisticated but less transparent parametrizations.
A different strategy has been to apply the low-energy expansion of the
effective Lagrangian for pion-pion scattering, studied by Gasser and
Leutwyler \cite{GL}.  This approach does not give manifestly unitary
scattering amplitudes, but it does give the correct crossing relations.
In particular, the three coefficients of order $s^2$ in the formulae
for the $a_I(s)$ are given in terms of two constants $L_1$, $L_2$.
Pion-pion scattering processes of new strong interactions have  been
studied in this parametrization in \cite{DandV,DHT,Pelaez,Kilian}.

\subsection{Strategies for Pion-Pion Scattering Experiments}

\begin{figure}[hbt]
\leavevmode
\begin{center}
\resizebox{!}{2.5in}{%
\includegraphics{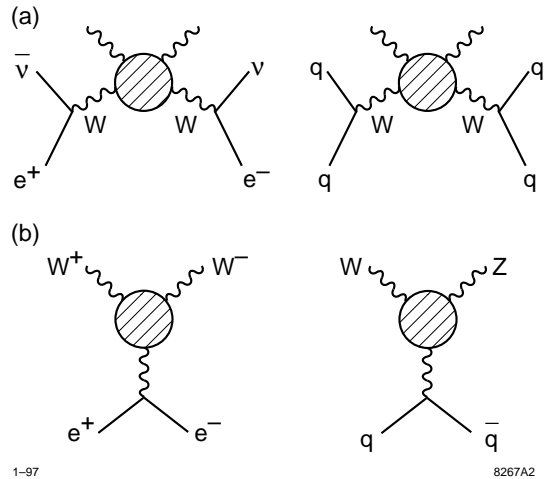}}
\end{center}
\caption{Processes which measure the pion-pion scattering amplitudes of 
       new strong interactions.}
\label{fig:WWscatt}
\end{figure}

There are two strategies for measuring the pion-pion scattering
amplitudes at colliders.  These are illustrated in
Figure~\ref{fig:WWscatt}. In the first of the processes shown,
longitudinally polarized $W$ bosons are radiated from incident
fermions, and these scatter to vector boson pairs at large transverse
momentum.  In the second process, the incident fermions annihilate
directly to vector boson pairs, which show the effects of new strong
interactions through the  production form factor.  The incident
fermions could be either quarks or leptons.  In either case, the
subprocess energy must be of order 1 TeV to see significant effects.

We will now make some general comments which provide theoretical
orientation on these two processes.  In Section IV, we will discuss
simulation studies of these two reactions and see what sensitivity they
can achieve in realistic settings.

The vector boson scattering process illustrated in
Figure~\ref{fig:WWscatt}(a) can in principle access all three of the
dominant channels of pion-pion scattering.  The channels are
distinguished by the identity of the final vector bosons:  $W^+W^+$ or
$W^-W^-$ couples only to $I$=2, $W^+Z^0$ to $I$=1,2, and $Z^0Z^0$ to
$I$=0,2.  The cross sections must be summed over the vector bosons
which appear in the initial state, though typically they are dominated
by channels with initial $W$'s.  This is especially true in $\ee$
reactions, since the probability for an electron to radiate a $Z$ is
about 1/10 of the probability to radiate a $W$.  In $pp$ reactions, all
of these processes occur in the same environment; at an $\ee$ machine,
experiments on the $I$=2 channel require dedicated operation of the
accelerator as an $e^-e^-$ collider.

The production of transversely polarized $W$ pairs is not influenced
significantly by the new strong interactions and so should be treated
as a background.  Often in the literature, the signal of new strong
interactions is expressed as the difference between the high-$p_T$
vector boson yield in the strong interaction model and the yield in the
minimal standard model with a Higgs boson of mass about 100 GeV.  The
latter assumption turns off the new strong interactions but retains the
production of transversely polarized boson pairs through standard model
gauge interactions.

Additional backgrounds come from processes of different topology that
lead to vector boson pairs or from final states that fake vector boson
pairs.  At the LHC, an important background reaction is the ubiquitous
$gg \to gg$, with the final jets broadened to the $W$ mass.  To
overcome this problem, it is typically required that vector bosons are
observed in their leptonic decay modes.  Even with this requirement,
processes such as $gg\to t\bar t$ and $gq \to W q$ must be removed
carefully with targeted cuts. At high energy $\ee$ colliders, the most
important  backgrounds come from photon-induced processes such as
$\gamma\gamma\to W^+W^-$ and $\gamma e \to W Z \nu$.  These are removed
by a forward electron veto and by requiring that the vector boson pair
has a transverse momentum of order $m_W$.  The kinematic requirements
of these cuts, and consideration of the overall rate, call for
observation of the $W$ and $Z$ in their hadronic decay modes.

In very favorable circumstances, a resonance of the new strong
interactions can produce a peak in the mass spectrum of vector boson
pairs.  However, the more typical situation is that it produces only a
shoulder or an enhancement over the prediction of a weak-coupling
model.  It is therefore a crucial experimental problem to understand
the standard model sources of vector boson pairs and other sources of
background {\em quantitatively}, so that these effects can be
subtracted accurately.

The fermion annihilation process illustrated in
Figure~\ref{fig:WWscatt}(b) is restricted to angular momentum $J$=1,
and therefore to the $I$=1 channel of pion-pion scattering.  We should
note, though, that in QCD this channel contains the $\rho$ meson and
thus has the strongest resonant effects at low energy.

We can parametrize the strong interaction effects on vector boson pair
production by a form factor $F_\pi(q^2)$, the pion form factor of the
new strong interactions.   The amplitude for $\ee$ or $q\bar q$
annihilation to new pions or longitudinally polarized $W$ and $Z$ pairs
is enhanced by the factor $F_\pi(s)$. The form factor satisfies
$F_\pi(0) = 1$ and should show an enhancement at the mass of any $I$=1
resonance.  If there is a strong $I$=1 resonance, the form of the
function will be described by a vector meson dominance parametrization
\beq
        F_\pi(s) =  {-M_1^2 + i \Gamma M_1\over s - M_1^2 + i \Gamma M_1}.
\eeq{Fpiform}
There is an additional piece of information about $F_\pi$ which is
known more rigorously.  Assuming only that $\pi \pi \to 4\pi$ processes
are unimportant at a given energy, the {\em phase} of $F_\pi(s)$ is
precisely the $I$=1 $\pi\pi$ phase shift $\delta_1(s)$.  Thus, even in
a nonresonant model, it is possible to detect the presence of new
strong interactions if one can measure the imaginary part of $F_\pi(s)$.

\begin{figure}[hbt]
\leavevmode
\begin{center}
\resizebox{2.5in}{!}{%
\includegraphics{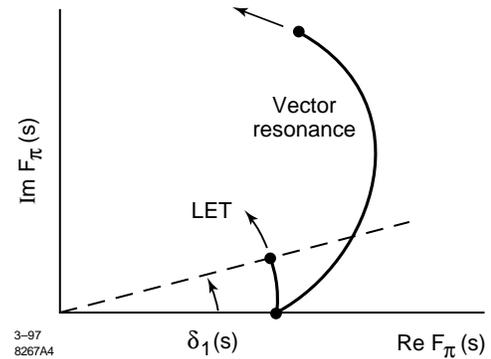}}
\end{center}
\caption{Dependence of $F_\pi(q^2)$ on energy, in models without and
with a new strong interaction resonance in the $I$=$J$=1 channel.}
\label{fig:FPi}
\end{figure}

In Figure~\ref{fig:FPi}, we show how the amplitude $F_\pi(s)$ moves in
the complex plane as $s$ increases, for the case of a nonresonant model
with phase shift $\delta_1(s)$.   We also show the motion of $F_\pi(s)$
for a resonant model in which, when the resonance is reached, the phase
shift goes through $\pi/2$.

At the  LHC, the main effect of the new pion form factor is to create
strong-interaction resonances in $q\bar q$ annihilation.  These can
show up as peaks in the $WZ$ mass distribution if the resonance mass is
sufficiently low.  In high-energy $\ee$ colliders, which are designed
to  measure the differential cross section for $\ee \to W^+W^-$ to 
percent accuracy, it is possible to measure the deviation of $F_\pi$
from 1 below the resonance or in a nonresonant case.  In particular, it
is possible to measure the phase of $F_\pi$ from the interference of
longitudinal and transverse $W$ pair production.

\section{Experimental Prospects for the Study of New Pion Interactions}

In this section, we will review simulation studies from the literature
and new studies that have been done at Snowmass that have explored the
experimental reach of future experiments on the new pion interactions.
These experiments are among the most difficult proposed for the next
generation of accelerators.  They require parton center-of-mass
energies above 1 TeV and luminosities sufficient to observe hard
scattering cross sections at these energies.  Thus, they tax the
highest energies and luminosities of future generations of colliders.

For this reason,  we will concentrate our attention on the LHC at full
design luminosity and a high-luminosity  electron-positron linear
collider with 1.5 TeV in the center-of-mass. Throughout this report, we
will use the shorthand `NLC' to refer to a next-generation $\ee$ collider.

The experimental studies that we will review in this section made a
variety of assumptions on the collider luminosity.  In this section,
whenever we present the results of a simulation, we will give  the size
of the data sample assumed. When we assemble and interpret these
results in Section VI, we will scale all results to data samples of 
100 fb$^{-1}$ (1 year at peak luminosity of $10^{34}$
cm$^{-2}$s$^{-1}$).  In all cases, we assume the full energy reach of
the machine (14 TeV for the LHC and 1.5 TeV for the NLC). In addition,
we assume 80\% electron polarization at the linear collider.  These
reference machine parameters are listed in Table~\ref{tab:parameters}.
At the end of  Section VI, we will comment briefly on what can be
learned from even higher energies and luminosities.

\begin{table}[hbt]
\begin{center}
\caption{Reference machine parameters, energy, luminosity, and
polarization, for  studies of new pion scattering.}
\label{tab:parameters}
\begin{tabular}{cccc}
\hline
\hline
  & CM Energy (TeV) & ${\cal L}$ (cm$^{-2}$s$^{-1}$) & $P$ \\
\hline
LHC & 14 & $10^{34}$ &  \\
NLC & 1.5 & $10^{34}$ & $P_e=80\%$\\
\hline
\hline
\end{tabular}
\end{center}
\end{table}

\subsection{Analysis Methods}
The strategies for studying these new strong interactions have been
discussed in Section IIIC.  The basic signals are (1) a peak in the
mass spectrum of vector boson pair production which is a
straightforward signal to extract, (2) a broad enhancement in the cross
section for boson boson scattering which involves an absolute cross
section measurement and a detailed understanding of all backgrounds,
and (3) a helicity analysis of $e^+e^- \rightarrow W^+W^-$ which allows
the measurement of the $I$=1 phase shift in that channel.

Many studies exist that probe the reach of the LHC and the NLC to
discover the new strong interactions.  Most studies rely on parton
level estimates for the signal and background processes.  For the LHC,
the most detailed studies are those of the detector collaborations
ATLAS and CMS~\cite{IanandJohn,cms,atlas}. The backgrounds are large,
and care has been taken to try to estimate the backgrounds
realistically.  Modern parton distribution functions tuned to Tevatron
data are used.  At
high luminosity, the LHC experiments expect 25 minimum bias events per
crossing. The effect of these events is included. Still, these analyses
do not involve full detector simulations.  Instead, GEANT simulations
of the momentum resolution, radiation losses and isolation effects in
the detectors are parametrized and then implemented with Gaussian
smearing.

At the NLC, where backgrounds and detector resolutions are much less of
a concern, most studies use reasonable parametrizations of detector
efficiencies and resolution, but do not involve detector simulation.
For the study of the reaction $e^+e^- \rightarrow W^+W^-$, four vectors
of stable particles emerging from the reactions are smeared assuming
Gaussian errors according to the parametrizations summarized in
\cite{NLCrep}. Studies of boson scattering processes such as $e^+e^-
\rightarrow \nu \bar{\nu} WW$ or $e^+e^- \rightarrow \nu \bar{\nu} ZZ$
are somewhat more idealized.

The goals for the simulation studies are to determine (1) whether we
can conclusively observe pion-pion scattering due to new strong
interactions, and (2) what can be determined about the structure and
resonances of the new strong interaction sector. In evaluating the
results we will present below, it is important to realize that, by the
time these experiments are done, the alternative explanation of
electroweak symmetry breaking through the vacuum expectation values of
light scalar fields will have been excluded at a high level of confidence.
We have discussed this point in  Section~IIB.

\subsection{Studies of New Strong Interactions  at the LHC}

At the  LHC, the main difficulty for experiments probing strong
symmetry breaking is backgrounds. In most studies, the two final state
gauge bosons are reconstructed in their ``gold plated" leptonic modes
only, although some studies have been able to demonstrate significant
signals in so called ``silver plated" modes where one gauge boson
decays leptonically and the other decays to 2 jets.
Table~\ref{tab:modesLHC} lists the detection modes and isospin channels
accessible at the LHC. (In our notation, $qq$ refers to a reaction of
quarks with the same or with different flavors.) The analyses suffer
from low statistics, in part because they select modes with small decay
branching ratios, but the high transverse momentum, isolated leptons
characteristic of the leptonic decay modes provide clean signatures for
gauge bosons in the final state.
 
\begin{table}[htb]
\begin{center}
\caption{Detection modes and isospin channels accessible at the LHC.}
\label{tab:modesLHC}
\begin{tabular}{cc}
\hline
\hline
Parton Level Process & Weak Isospin \\
\hline
$qq \to qqZZ$ & 0, 2 \\
$qq \to qqW^+W^-$ & 0, 1, 2 \\
$qq \to qqWZ$ & 1, 2 \\
$qq \to qqW^\pm W^\pm$ & 2 \\
$q \bar q  \to ZZ$ & 0,2 \\
$q \bar q \to Z\gamma$ & 0 \\
$q \bar q \to ZW$ & 1 \\
\hline
\hline
\end{tabular}
\end{center}
\end{table}
 
\subsubsection{Resonance Searches}

Both ATLAS and CMS have studied the ability of their experiments to
discover resonances in diboson production due to  new strong
interactions.  The ATLAS technical proposal  \cite{atlas} describes a
search for evidence of a technirho ($\rho_T$) decaying to $WZ$ or a
techniomega ($\omega_T$) decaying to $Z\gamma$.

In the $I$=1 channel, they estimate that, for a 1 TeV $\rho_T$, the
signal production cross section is 40 fb.  This comes from a 
combination of direct production ($q q \rightarrow WZ$) and boson-boson
scattering ($q q \rightarrow q q W Z$), though the annihilation channel
dominates. The experimental signature is three isolated, high
transverse momentum leptons in the final state, together  with
significant missing energy. In the study, the missing transverse energy
was computed after energy smearing, using the momentum vectors of all
particles observable in the detector. A $W$ mass constraint was then
applied to the lepton-neutrino pair in order to reconstruct the total
invariant mass of the $WZ$ system.

\begin{figure*}[tbt]
\leavevmode
\begin{center}
\resizebox{!}{4.0in}{%
\includegraphics{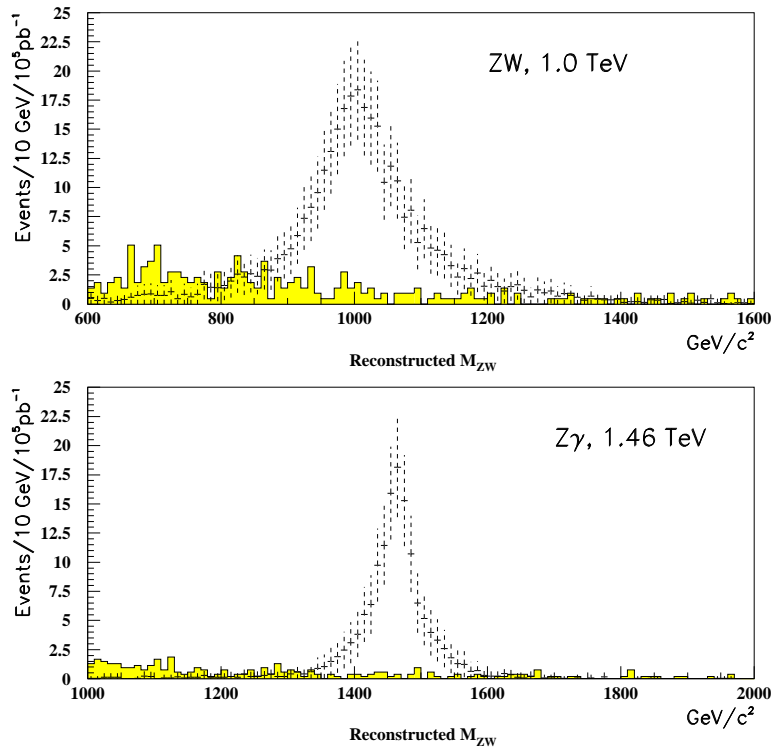}}
\end{center}
\caption[*]{Reconstructed masses for high mass resonances decaying into
gauge boson pairs, from 
\protect\cite{atlas}: (a) Signal of a 1 TeV $\rho_T$
decaying to $WZ$ and subsequently to three leptons  (b) Signal of a
1.46 TeV $\omega_T$ decaying to $Z\gamma$ with the $Z$ then decaying to
two leptons.}
\label{fig:lhcres}
\end{figure*}

The dominant background is $t\bar{t}$ production with one dilepton
combination having a mass close to the $Z$ mass.  This background is
effectively removed by the isolation cuts on the leptons and the $W$
mass constraint described above.  The other backgrounds are reactions
that produce transversely polarized $WZ$ pairs, continuum $q \bar{q}
\rightarrow WZ$ and boson-boson scattering through $q q \rightarrow q q
W Z$.    Note that because the cross section is dominated by the
annihilation channel, there is no advantage in tagging forward jets to
select the boson-boson scattering reaction.

The reconstructed diboson mass for a 1 TeV $\rho_T$ decaying to $WZ$,
for 100 fb$^{-1}$ of data at the LHC, is shown in 
Figure~\ref{fig:lhcres}(a).  The backgrounds remaining after all cuts
are small and the resonance is sufficiently narrow that the signal
would be straight\-forward to observe. This analysis can be extended to
search for vector resonances at LHC up to a mass of 1.6 TeV with 100
fb$^{-1}$ of data \cite{atlas,wulz}.

The search for an $\omega_T$ resonance is quite similar. As we have
noted in Section IIIB, the $\omega_T$ is not produced in boson-boson
scattering; also, it is a model-dependent question whether it couples
to $q \bar q$ annihilation. If it does couple, it can produce a
dramatic effect, since the $\omega_T$ can decay to $\gamma Z^0$,  which
is directly reconstructed if the $Z$ decays leptonically.   The
dominant backgrounds are continuum $Z\gamma$ production and $Z +$ jet
production where the jet fakes an isolated photon.  The production
cross section for a 1.46 TeV $\omega_T$ is expected to be approximately
50 fb. Figure~\ref{fig:lhcres}(b) shows the reconstructed $Z\gamma$
mass distribution for the expected signal in a 100 fb$^{-1}$ data set. 
This search is sensitive to $\omega_T$ up to approximately 2 TeV.

\subsubsection{Absolute Cross Section Measurements}

If the resonances in the new strongly coupled world are too massive or
too broad, we will not  be able to see them directly as peaks in a
diboson mass spectrum. In that case, we can only look for the effects
of the new interactions as event excesses or cross section enhancements
in the various boson-boson scattering channels. The largest
enhancements will appear in the isospin channel corresponding to the
quantum numbers of the strongest resonance.  At a certain point,
however, our sensitivity to the mass of the object dominating the
scattering channel will be lost.   Then we may still be able to
establish the existence of new strong interactions, but we will be
unable to distinguish anything about their detailed structure.
 
At the LHC, the  most sensitive channel for  probing the high mass
structure of the new strong interaction is $qq\to qqW^\pm W^\pm$.  This
channel minimizes the backgrounds from pair-production of transversely
polarized $W$ bosons, since the  $W^\pm W^\pm$ final state cannot
result from an annihilation process, and it minimizes the  background
from $t \bar t$ production leading to hard isolated leptons.

The experimental signal is the observation of two isolated same-sign
leptons in the central detector, with large dilepton invariant mass.
The dominant backgrounds come  from electroweak bremsstrahlung
processes, gluon exchange, $W + t\bar{t}$ production, $WZ$ and $ZZ$
final states, and $t\bar{t}$ production. In the ATLAS study of this
decay mode, the dominant background processes from $t\bar{t}$, $WZ$ and
$ZZ$ were generated using PYTHIA 5.7 and the other backgrounds were
simulated at the parton level.  Detector acceptance and resolution for
leptons and jets were included in the simulation.

The signal could only be observed above background if the analyses made
cuts on the distinctive topology of the boson scattering process. 
Signal events have two jets in the forward regions of the detector from
the scattered quarks, but there is not much jet activity in the central
detector if one has required both $W$ bosons to decay leptonically. 
The $t\bar{t}$ background is greatly reduced by rejecting events with a
jet of significant energy in the central detector, and all backgrounds
not originating from boson fusion are reduced by requiring the tagging
of two  jets, one in each of the forward detector regions.  The
dominant background remaining after all cuts is transverse $W$ pair
production.

\begin{figure}[htb]
\leavevmode
\begin{center}
\resizebox{!}{3.0in}{%
\includegraphics{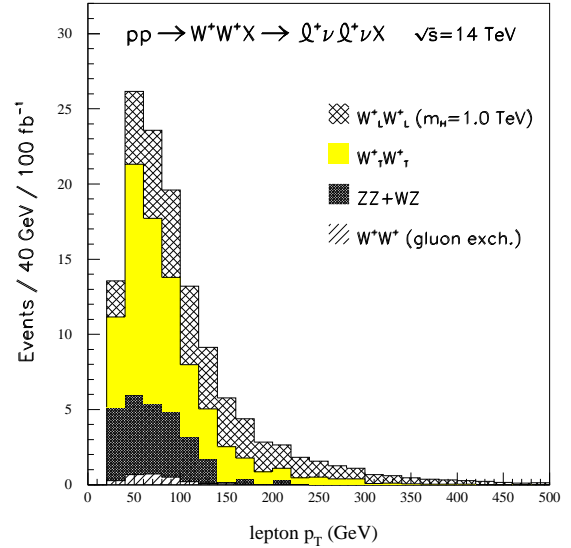}}
\end{center}
\caption[*]{Expected numbers of $W^+W^+ \rightarrow (l\nu)(l\nu)$ signal 
and background events after all cuts for a 100 fb$^{-1}$ data
sample at the  LHC, from \protect\cite{atlas}.  The signal corresponds to a
1 TeV Higgs.}
\label{fig:nonres}
\end{figure}

The event rates in this channel are low and the signal to background
ratio does not change much as a function of lepton transverse energy.
Figure~\ref{fig:nonres} shows the signal expected for a model in which
a 1 TeV $I$=$J$=0 particle (a massive Higgs scalar) is exchanged in the
$t$-channel. The review  \cite{Golden} provides a comprehensive survey
of this channel for a variety of hypotheses of the source of the strong
symmetry breaking (scalar or vector resonances with a variety of
masses).  In all cases, there is no resonance shape to distinguish
signal from background and the signal can only be extracted if the
absolute magnitude of the background is known from other sources. 
However, if one does understand the backgrounds \cite{darian} and can
accurately model the jet tagging and jet vetoing in the high luminosity
environment of the LHC, studies show that the $qq\to qq'W^\pm W^\pm$
channel is sensitive to strong electroweak symmetry breaking even in 
the LET limit.   No matter how high one places the masses of
resonances, there will be a significant excess of events over
background in this channel. In other words, this channel will discover
the new strong interactions if they are there.

The yields in this channel, in the LET limit, for a 100 fb$^{-1}$ data
sample are listed in the general summary of the results of simulation
studies which we present in Table IV  of Section VI.
 
\subsection{Studies of New Strong Interactions  at the NLC}

At the NLC, the available channels for studying strong symmetry
breaking are limited by rate rather than backgrounds.  The boson-boson
scattering processes that can be studied are listed in
Table~\ref{tab:bosonprocess}. All studies use the 2 jet decays of the
gauge bosons to take advantage of the largest accessible branching
ratios. The $I$=2 final state $W^-W^-$ can also be studied  at a
linear collider; however, this  requires a dedicated experiment of
several years' duration with $e^-e^-$ collisions.  It is interesting
that this information is potentially available, but we will not
consider this process further here.

\begin{table}[bht]
\begin{center}
\caption{Detection modes and isospin channels accessible at the NLC.}
\label{tab:bosonprocess}
\begin{tabular}{cc}
\hline
\hline
Parton Level Process & Weak Isospin \\
\hline
$e^+e^- \to \nu\bar \nu ZZ$ & 0, 2 \\
$e^+e^- \to \nu\bar \nu W^+W^-$ & 0, 1, 2 \\
$e^+e^- \to W^+W^-$ & 1 \\
\hline
\hline
\end{tabular}
\end{center}
\end{table}

Another tool available at NLC is the annihilation channel $e^+e^- \to
W^+W^-$.  The rates are high enough and backgrounds low enough that one
can do a helicity analysis by reconstructing the  production and decay
angles of the $W$'s in the final state. This allows one to separate the
longitudinal and transverse contributions to the final state, and to
measure the new pion form factor discussed in Section IIIC.

\subsubsection{Boson-Boson Scattering at the  NLC}

A study of the boson scattering processes at an $e^+e^-$ linear
collider, $e^+e^- \rightarrow \nu \bar{\nu} WW$ and $e^+e^- \rightarrow
\nu \bar{\nu} ZZ$ is described in \cite{Barger}
 and  reviewed in a contribution
\cite{barksnow} to these proceedings. Here we will summarize only the
essential features of the analysis.  The gauge bosons are each
reconstructed in the dijet final state.  With realistic jet energy
resolutions, the $W$ and $Z$ bosons cannot be discriminated on an
event-by-event basis, but they can be separated statistically.  The
dominant backgrounds are dibosons produced from $e^+e^-$ annihilation,
$e^+e^- \rightarrow e^+e^-W^+W^-$, and $\ee \to e \nu W Z$, with
misidentification of $WZ$ events as either $WW$ or $ZZ$ events due to
jet energy smearing.  The study \cite{Barger} does not include detector
resolution or efficiencies except in the important area of jet energy
resolution where $\Delta E_j/E_j = 50\%/\sqrt{E_j} \oplus 2\%$ is
assumed.

The annihilation background is removed by a cut requiring significant
missing mass in the event. One first requires $W^+W^-$ events  to have
large $WW$ invariant mass and a large angle with respect to the beam
axis ($|\cos\theta_W| < 0.8$).  The dominant backgrounds then come from
$e^+e^- \rightarrow e^+e^-W^+W^-$  and $\ee \to e \nu W Z$, due to 
intermediate states with virtual photons radiated from the electron
lines. These backgrounds are suppressed by the requirement that the
transverse momentum of the $WW$ system be large (50 GeV $< p_T(WW) <$
300 GeV), and by a veto on events with hard forward electrons. For the
veto, it  is assumed that electrons with $E_e > 50$ GeV can be tagged
for $|\cos\theta_e| < 0.99$ \cite{NLCrep}. With this series of cuts,
the strong symmetry breaking signals are observable over background.

\begin{figure}[bht]
\leavevmode 
\begin{center}
\resizebox{!}{4.50in}{%
\includegraphics{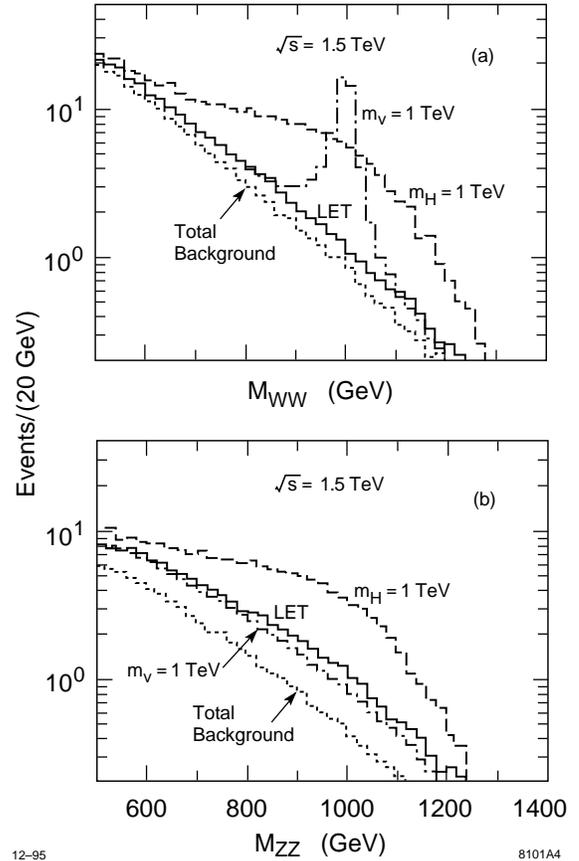}}
\end{center}
\caption{Expected numbers of $W^+W^-, ZZ \to (jj)(jj)$ signal and
background events after all cuts for a 200 fb$^{-1}$ unpolarized data
sample at the NLC at 1.5 TeV, from \protect\cite{Barger}:
 (a) $e^+e^- \rightarrow
\nu \bar\nu W^+ W^-$, (b) $e^+e^- \rightarrow \nu \bar\nu Z Z$.  The
dotted histogram shows the total standard model background including
misidentifications.  The solid histogram shows signal plus background
for the LET.}
\label{fig:nlcbb}
\end{figure}

The resulting $WW$ and $ZZ$ signals, for a 200 fb$^{-1}$ data sample at
a 1.5 TeV NLC, are shown in Figure~\ref{fig:nlcbb} as a function of the
diboson invariant mass. As was the case at the LHC, the signal to
background ratio does not change significantly as a function of this
mass, and there is no resonance shape to distinguish signal from
background unless the resonance masses are very low (or order 1 TeV).
Thus, again, the signal can only be extracted if the absolute magnitude
of the background is known from other sources.  However, if the
background can be accurately modeled, the study of \cite{Barger} shows
that the $e^+e^- \rightarrow \nu \bar{\nu} ZZ$ channel is sensitive to
strong electroweak symmetry breaking to the LET limit.

\subsubsection{Helicity Analysis of $e^+e^- \rightarrow W^+W^-$ at the NLC}

As was discussed in Section IIIC, the annihilation channel $e^+e^-
\rightarrow W^+W^-$ can be used to probe the new strong interactions
because final state rescattering provides information about the $J$=1
partial wave in the process $W^+_LW^-_L \rightarrow W^+_LW^-_L$.  The
strategy of the analysis is to use the decay angular distribution of
the $W$'s to measure the final state $W$ polarization, and then to
extract the real and imaginary parts of the form factor from the
contribution of  $e^+e^- \rightarrow W^+_LW^-_L$.

\begin{figure}[hbt]
\leavevmode
\begin{center}
\resizebox{!}{1.6in}{%
\includegraphics{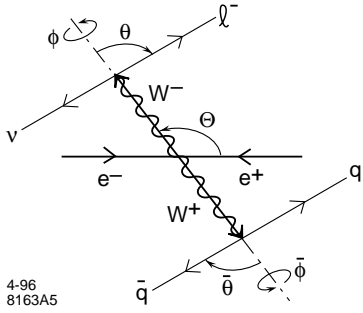}}
\end{center}
\caption{Production and decay angles in $e^+e^- \rightarrow W^+ W^-$,
   used for the extraction of the new pion form factor $F_\pi(s)$.}
\label{fig:helicity}
\end{figure}

The final state topology with one $W$ decaying hadronically and the
other decaying leptonically is best for the analysis.  The experimental
observables for the polarization analysis are illustrated in
Figure~\ref{fig:helicity}.  They are the $W^-$ production angle
($\Theta$), the polar and azimuthal angles of the lepton in the $W$
rest frame  ($\theta, \phi$), and the polar and azimuthal angles of the
quark jets in the $W$ rest frame ($\bar{\theta}, \bar{\phi}$).  No
quark flavor tagging is assumed so that the two quark directions are
averaged over. Events are generated by a Monte Carlo program that
retains the full spin correlations through the $W$ decay \cite{barklow}.

Detector effects are simulated by smearing the four vectors of the
final state particles.  The analysis uses only two cuts to select
events. $|\cos\Theta| < 0.8$ ensures that the event is within the
detector volume and the $W$ invariant masses are required to be within
a few GeV of the known $W$ mass, where the mass of the leptonically
decaying $W$ is reconstructed by using 4-momentum conservation in the
event to solve for the 4 vector of the undetected neutrino.  The cuts
yield a sample that is 98\% pure with 36\% efficiency (not including
the $W$ branching ratios).

\begin{figure}[htb]
\leavevmode
\begin{center}
\resizebox{!}{2.8in}{%
\includegraphics{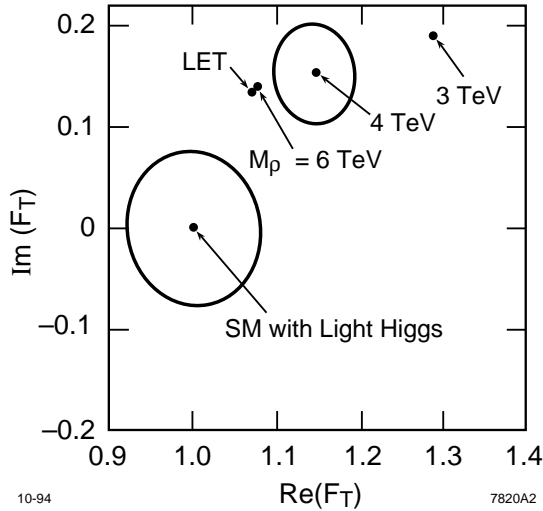}}
\end{center}
\caption[*]{Determination of the new pion form factor $F_\pi(s)$ at the
NLC at 1.5 TeV with an unpolarized data sample of 200 fb$^{-1}$, from
\protect\cite{barklow}.
 The results are compared to a model with a high-mass $\rho_T$
and LET behavior as this mass goes to infinity.    The contour about
the light Higgs value is a 95\% confidence contour; that about the
point $m_{\rho} = 4$ TeV is a 68\% confidence contour.}
\label{fig:nlcWWres}
\end{figure}

The unpolarized differential cross section for $e^+e^- \rightarrow W^+
W^-$ would be directly sensitive to relatively light vector resonances
without any polarization analysis.   For example, a $\rho_T$ resonance
at the collider center-of-mass energy increases the polarization-summed
differential cross section at 90$^\circ$ by a factor of 4.  A more
challenging question is that of the sensitivity to higher-mass $I$=1
resonances, or to the LET limit. Figure~\ref{fig:nlcWWres} shows two
illustrative determinations of $F_\pi(s)$, through a maximum likelihood
fit to simulation data, for the NLC at 1.5 TeV with an unpolarized data
sample of 200 fb$^{-1}$.  These determinations are compared to an
explicit model with a high-mass vector resonance. The points illustrate
the general behavior shown in Figure~\ref{fig:FPi}. The LET limit is
distinguishable from the case of no new strong interactions at the 4.6
$\sigma$ level.  This translates to a 4.3 $\sigma$ significance for a 
100 fb$^{-1}$ data sample with 80\% polarization.

\section{The Top Quark Coupling to New Pions}

In the previous two sections, we have discussed how the new strong
interactions responsible for electroweak symmetry breaking are probed
by the study of $WW$ scattering, $W$ pair production, and the $W$ gauge
couplings.  All of these experiments refer only to the property that
the new strong interactions spontaneously break $SU(2)\times U(1)$. But
the mechanism of electroweak symmetry breaking has one more important
task, to give mass to quarks and leptons.  It is a very important part
of the general program of experiments on the new strong interactions to
probe how this sector couples to fermions.

A particular mystery surrounds the mass of the top quark.  This quark
must be especially tightly coupled to the new sector.  It is tempting
to assume that the top quark actually plays an essential role in the
mechanism of $SU(2)\times U(1)$ symmetry breaking.  It is also possible
that new strong interactions of the top quark are shared, to some
extent, by the other third generation fermions $b$ and $\tau$. On the
other hand, it is equally well possible to have a sector of new strong
interactions in which the top quark does not play an essential role.

These different possibilities are reflected in formal considerations
for the top quark scattering amplitudes. The low-energy theorem for the
process  $t \bar t \to \pi \pi$, has a steep rise with $s$ similar to
that which we discussed for $\pi\pi\to \pi\pi$ in  \leqn{aasym}
\cite{CFH,TApp,Goldent},
\beq
           a_0 = m_t \sqrt{s}/16 \pi f_\pi^2  + {\cal O}(s^2) \ , 
\eeq{tasym}
This formula implies the unitarity bound
\beq
    I = 0:\ \sqrt{s}<  16  \hbox{\rm TeV} \ . 
\eeq{tlimits}    
It is an interesting question at what energy the expression
\leqn{tasym} turns over to respect unitarity.  It could be unitarized
by the resonances of the $\pi\pi$ system, at the much lower energies
suggested in \leqn{ulimits}.  In this case, the top quark would couple
to these resonances as a perturbation, with an amplitude proportional
to  $(m_t/4\pi f_\pi)$. Alternatively, the $t \bar t$ system could
couple to new resonances, which might appear either at very high
energies or near the TeV scale.

In Section VIII, we will discuss some specific models of fermion mass
generation.  Within these particular models, there are definite
signatures of new physics associated with heavy quarks which can be
searched for experimentally.  In particular, these models suggest the
existence of exotic particles decaying to heavy flavors. We will review
searches for these particles in Section IX.

Here we address the question of whether the $t\bar t - \pi\pi$
amplitude can be directly studied experimentally.  The most direct way
to do this would be to create top quark pairs by pion-pion scattering. 
This involves the study of the process $W^+W^- \to t \bar t$. The
analysis of this process is similar to that of $WW$ scattering. The $W$
bosons are produced by radiation from external fermions lines. There is
a contribution from transversely polarized $W$ bosons scattering
through the usual standard model interactions.  The more interesting
contribution from longitudinally polarized $W$ boson may contain
enhancements or resonances due to the new strong interactions.

\begin{figure}[bht]
\leavevmode
\begin{center}
\resizebox{1.6in}{!}{%
\includegraphics{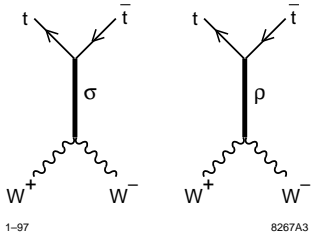}}
\end{center}
\caption[*]{Contributions to $W^+W^-\to t \bar t$ from new strong
interactions. In this figure, $\sigma$ is an $I$=0 resonance; $\rho$ is
an $I$=1 resonance.}
\label{fig:WWtt}
\end{figure}

The two most important contributions of the latter type are shown in
Figure~\ref{fig:WWtt}.  The coupling of the top quarks and $W$ bosons
to a scalar resonance should be proportional to the masses these
particles acquire from the Higgs mechanism.  Thus, the first diagram of
Figure~\ref{fig:WWtt} has an amplitude
\beq
                {m_t \sqrt{s}\over f_\pi^2} {M_0^2 \over s - M_0^2}\ .
\eeq{ttspinzero}
This amplitude becomes large for large $M_0$;	in the limit $M_0 \to
\infty$, it violates unitarity at high energy.  This process is
helicity-flip, producing the final states $t_L {\bar t}_L$ and $t_R
{\bar t}_R$ predominantly for $s\gg m_t^2$.  The second diagram of 
Figure~\ref{fig:WWtt} involves the lowest $I$=1  ($\rho_T$) resonance. 
This resonance has a more model-dependent coupling for which we have
introduced a parameter $\eta$.  The value of the amplitude is
\beq
                {\eta m_t M_1\over f_\pi^2} {s\over s - M_1^2}\ .
\eeq{ttspinone}
This process is helicity-conserving, leading to the final states
$t_L {\bar t}_R$ and $t_R {\bar t}_L$ at high energy.

\begin{figure*}[t]
\leavevmode
\begin{center}
\resizebox{!}{3.5in}{%
\includegraphics{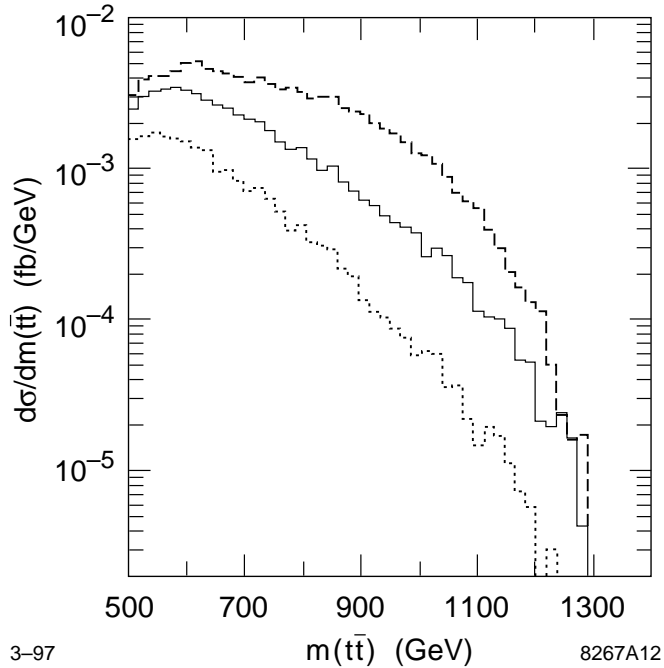}}
\end{center}
\caption{Production cross section for the process $WW\to t\bar t$,
in the 6-jet topology after all cuts,
as a function of the $t\bar t$ mass,
for a 1.5 TeV NLC with 100\% electron polarization, from 
\protect\cite{barksnow}.
  The dotted curve is the standard model
 background; the dashed curve is the effect of a scalar resonance at 
1 TeV; the solid curve is the LET prediction.}       
\label{fig:differential}
\end{figure*}

Because these two processes lead to different final polarization states
of the top quark, it is possible, in principle, to separate them.  Each
of the two processes gives separate information about the relation of
the top quark to the mechanism of $SU(2)\times U(1)$ breaking.  For the
first process, we can ask, is the value of the resonance mass $M_0$ the
same as that observed in $I$=0 $WW$ scattering?  We will describe below
a model in which the reaction $WW\to t\bar t$ has a lower resonance
position, corresponding to a mechanism for generating the top quark
mass which is distinct from the mechanism for generating the $W$ mass. 
For the second process, we can ask, is $\eta$ close to 1, signaling a
significant coupling of the top quark to the $I$=1 resonance, or is it
very small?  Thus, even before looking into the details of models, the
measurement of this process will give us guidance as the the origin of
the top quark mass.

The observation of $WW\to t\bar t$ at the LHC seems very difficult due
to the overwhelming background from $gg\to t\bar t$.  However, it
appears to be feasible to study this process at the NLC.  A simulation
study has been done of the reaction $\ee\to \nu\bar \nu t\bar t$, which
contains  $WW\to t\bar t$ as a subprocess, assuming contributions from
an $I$=0  resonance only \cite{barksnow}.  The standard model
predictions for this process have been presented in \cite{GintGod}. As
in the boson-boson scattering studies at the NLC, it is advantageous to
reconstruct the top quarks in their hadronic decay modes. The
experimental signature  is either 6 jets (two of which are potentially
$b$ tagged) or 4 jets and a lepton (again with potentially two $b$ tags).
The final jets (and leptons) are kinematically fit to form a $t\bar t$
system. The dominant background is two-photon production of $t \bar t$
pairs through the process $e^+e^- \to e^+e^-t \bar t$. This background
can be suppressed by requiring  transverse momentum of the $t\bar t$
system and a forward electron veto, just as in  the analysis of boson
scattering processes.
 
The cross section for the 6-jet topology at the NLC at 1.5 TeV, after all
cuts have been applied, is shown in Figure~\ref{fig:differential}.
 The highest curve
is the prediction of a model with an $I$=0 resonance at 1 TeV.  Even in
this case, there is no clear resonance shape but only an enhancement
above the background.  However, the predicted enhancement is
significant not only for this example but also for the LET limit.  The 
event yield in that case is given in Table~\ref{tab:sumnum}.

\begin{table*}[t]
\begin{center}
\caption{Summary of yields from the LHC and NLC studies described in 
     Sections IV and V, for the LET limit, 
     scaled to 100 fb$^{-1}$ data samples.}           
\label{tab:sumnum}
\begin{tabular}{cccccc}
\hline
\hline
Process & $N$(signal) & $N$(bkgd) & $S/\sqrt{B}$ & ref.  \\
\hline
$qq' \to qqW^\pm W^\pm $  & 39 & 76 & 4.5 & \protect\cite{atlas} \\
$qq \to qqZ^0  Z^0 $ & 10.4   & 3.3    & 5.7  & \protect\cite{bagger}   \\
$e^+e^- \to \nu \bar \nu Z^0Z^0$  & 41 & 50 & 5.7 & \protect\cite{Barger} \\
$e^+e^- \to \nu \bar \nu t \bar t$  & 53  & 36 & 8.8 &
\protect\cite{barksnow}  \\
\hline
\hline
\end{tabular}
\end{center}
\end{table*}

\section{Summary of Experimental Prospects at LHC and NLC}

Now that we have reviewed the major simulation studies of vector boson
scattering processes at future colliders, we must try to place the
results in a coherent picture.  What do these results mean
qualitatively?  What will we really learn about the new strong
interactions from the LHC?  What will the NLC add to our understanding?

\subsection{Existence of New Strong Interactions}

To begin, let us summarize the results of the previous sections. From
the studies of the vector boson scattering, we have seen that the LHC
can establish the presence of new strong interactions in the  channel
$qq\to qq W^\pm W^\pm$ even at the level of the LET predictions. The
estimates in the literature of the experimental significance of such a
measurement range from 3.5 to 7 sigma. The most conservative numbers
are from the ATLAS Technical Proposal \cite{atlas} using double jet
tagging. The NLC can independently establish the presence of new strong
interactions in the $\ee \to \nu \bar\nu Z^0 Z^0$, and $e^+e^-\to\nu\nu
t \bar t$ channels. A summary of the yields reported in  the NLC and
LHC studies is given in Table~\ref{tab:sumnum}. In this table, we have
scaled the simulation results presented previously to 100  fb$^{-1}$
data samples, with 80\% polarization for the $\ee$ reactions. All of
the results that we will present in this section assume the statistics
of samples of this size.

In addition, we have discussed the probe of the new strong interactions
given by the search for $I$=1 resonances, or, equivalently, in the
study of the new pion form factor.  We have seen that the LHC will
detect $I$=1 $WZ$ resonances up to masses of 1.6 TeV.  The study of
$\ee\to W^+W^-$ at the NLC will be sensitive to the form factor effect
even in the LET limit.

Beyond these pieces of direct evidence, the exclusion of  a light Higgs
boson coupling to $ZZ$ would force us to consider new strong
interactions as the only alternative to explain electroweak symmetry
breaking.

\subsection{Sensitivity to  New Resonances}

Demonstrating the existence of new strong interactions, however, is
only the first step.  In order to build and test models of these new
interactions, we will need to obtain the basic experimental information
on their structure. In principle, one could imagine going to very high
energy and mapping out the structure of resonances of the new strong
interactions, as we have done in QCD. However, this is not a realistic
goal for the next generation of colliders.  We found it interesting to
ask a more modest question: Can the colliders of the next generation
identify the lowest-lying resonances of the new strong interactions and
measure their masses? For example, can the LHC and the NLC distinguish
a model in which the new strong interactions are described by quark
model, with a strong $I$=$J$=1  resonance, from one resembling a
strongly coupled Higgs model with a strong $I$=$J$=0 resonance and
little activity in the $I$=1 partial wave?

The best way to address this question would be to perform new
simulations using a variety of possible strong interaction models.  We
have not done that in this study. However, we believe that it is
possible to reinterpret the results of the simulation studies we have
reviewed in Sections IV and V to give a first quantitative answer to
this question.  We will now discuss three strategies to this conclusion.

\subsubsection{Comparison of Models}

The first approach we follow is to ask whether specific models studied
by simulation in the literature could be distinguished on the basis of
the simulation results.  A particularly interesting study for this
purpose is that of \cite{bagger}.  This analysis compared the signal
strength at the LHC in a variety of channels for five specific models
of the new strong interactions: the standard model with a 1 TeV Higgs
boson, a model with a scalar resonance at 1 TeV, a model with a vector
resonance at 1 TeV, a model with a vector resonance at 2.5 TeV, and a
nonresonant model.  The first two models differed primarily in the
width of the scalar resonance, which was taken to be 0.49 TeV and 0.35
TeV, respectively, in the two cases.  In the fifth model, the pion-pion
scattering amplitudes were given by the LET with K-matrix
unitarization. For each model, the signal strength was computed for a
fixed set of experimental cuts.  Only the gold-plated modes were
considered, and so the yields are rather small.

Given this data, one can ask the question,  if these yields are
fluctuated statistically, to what extent could the results for one
model be fit by one of the other models?  This analysis is described in
detail in a separate paper in these proceedings \cite{kilgore}. The
result is  that the first two  models are clearly distinguished from
the others.  Further,  these two models give  a poor fit to one
another, with a $\chi^2$ per degree of freedom greater than 2 for the
wrong choice. The models with vector resonances were not distinguished
from the LET model. However, this is not a surprise, since these are
distinguished from the LET primarily by the search for resonance
production via $q \bar q \to W Z$, and this reaction is not selected by
the cuts of \cite{bagger}.  We have seen that, when the mass of the
vector resonance is as low as 1 TeV, it is a prominent feature that
would distinguish this model.
 
\subsubsection{Analysis  of Yields}

It is possible to go a step further with the data from the simulation
studies. To the extent that an experiment can distinguish the LET
prediction from the case without new strong interactions, it can also
be sensitive to an enhancement of the same signal due to new strong
interaction resonances.  As a convenient basis for analysis, we might
parametrize the new pion scattering amplitudes with simple formulae
depending on resonance masses, as we discussed in Section III.  Then we
can ask, how well can the parameters in these formulae be measured in
various reactions.

We can give a rough but quantitative answer to this question for any
process in which we understand the accuracy with which the signal of
new strong interactions is measured.  Thus, we can estimate the
sensitivity to resonances of experiments on boson-boson scattering and
on the new pion form factor.

We emphasize that, in this analysis, we are not insisting that the
resonance be detected as a peak in the $WW$ scattering cross section.
We are asking only whether the resonance can be detected as an excess
of events that may be associated with a new strong interaction
resonance. The resonance peak may be difficult to detect either because
the resonance is very broad (as in the case $m_H = 1$~TeV in
Figure~\ref{fig:nlcbb}) or because it is at an energy higher than the
parton-parton center-of-mass energy (the typical situation for an $I$=1
resonance which gives a small enhancement to  the new pion form
factor). If there are two important resonances in a scattering channel,
the effective mass parameter will be a combination of the two masses. 
It is also possible that two resonances in a single channel could
destructively interfere at low energy and produce no observable effect.
Nevertheless, we feel it is interesting to ask how well the data from
future colliders will determine one parameter for each partial wave
beyond the model-independent scattering lengths given in \leqn{Avalues}.

\begin{figure}[hbt]
\leavevmode
\begin{center}
\resizebox{!}{1.8in}{%
\includegraphics{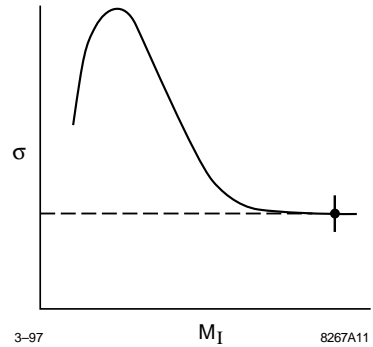}}
\end{center}
\caption[*]{Cross section for the process $\pi\pi\to \pi\pi$ in a given 
   partial wave, at fixed $\pi\pi$ center-of-mass energy, as a function 
     of the mass $M$ of a resonance in this channel.}
\label{fig:peak}
\end{figure}

\begin{table*}[ht]
\begin{center}
\caption{LHC and NLC sensitivity to resonances in the new strong
interactions. `Reach' gives the value of the resonance mass
corresponding to an enhancement of the cross section for boson-boson
scattering at the 95\% confidence level obtained in Section VIB2.
`Sample' gives a representative set of errors for the determination of
a resonance mass from this enhancement. `Eff. $\L$ Reach' gives the
estimate of the resonance mass for a 95\% confidence level enhancement
obtained in Section VIB3. All of these estimates are based on simple
parametrizations in which a single resonance dominates the scattering
cross section. A more complete explanation of the assumptions used is
given in the text.}
\label{tab:Peskin}
\begin{tabular}{cccccc}
\hline
\hline
Machine & Parton Level Process & I & Reach & Sample & Eff. $\L$ Reach \\ 
\hline \\
LHC & $qq' \to qq'ZZ$ & 0 & 1600 & $1500^{+100}_{-70}$& 1500 \\ \\
LHC & $q \bar q \to WZ$ & 1 & 1600 & $1550^{+50}_{-50}$ & \\ \\
LHC & $qq' \to qq'W^+W^+$ & 2 & 1950 & $2000^{+250}_{-200}$&  \\ \\
NLC & $e^+e^- \to \nu \bar \nu ZZ$ & 0 & 1800 & $1600^{+180}_{-120}$&
 2000 \\ \\
NLC & $e^+e^- \to \nu \bar \nu t \bar t$ & 0 & 1600 & $1500^{+450}_{-160}$&
\\ \\
NLC & $e^+e^- \to W^+W^-$ & 1 & 4000 & $3000^{+180}_{-150}$ \\ \\
\hline
\hline
\end{tabular}
\end{center}
\end{table*}
 
Consider first the case of boson-boson scattering.  In
Figure~\ref{fig:peak}, we show a typical $\pi\pi$ scattering cross
section at fixed $\pi\pi$ center-of-mass energy, as a function of the
resonance mass $M_I$ defined in \leqn{standarddelta}, with  $I$ taken
to be the dominant partial wave for this reaction. The cross section
peaks at the value of $M_I$ equal to the center-of-mass energy used. If
$M_I$ is much larger than the center-of-mass energy, the resonance
recedes and the cross section decreases.  As $M_I\to \infty$, the cross
section approaches a  limiting value; this is precisely the LET model.

For each simulation study in Table~\ref{tab:sumnum}, we have the
expected number of signal and background events for the case in which
the $\pi\pi$ scattering process is governed by the LET model. From the
quoted yields in each simulation, and assuming that the dominant source
of error is the statistical fluctuation of the small number of signal
and background events, we can compute the accuracy with which the
predicted LET cross section is measured.  Then, assuming that we know
the shape of the cross section curve with $M_I$, we can ask: (1) At
what value of $M_I$ does an event excess appear at the 95\% confidence
level? (2) At a given value of $M_I$ at the upper end of the observable
range, what is the uncertainty in the determination of $M_I$ from the
event excess?

To answer these questions, we made several further simplifications.  We
did not redo the simulations to compute the dependence of the cross
sections on $M_I$.  Rather, we used the dependence on $M_I$ of
scattering cross sections at the $\pi\pi$ center-of-mass energy equal
to the typical value 1 TeV. For the reactions $WW\to ZZ$, which receive
contributions from $I$=0,2, we assumed that the resonance producing the
event excess was in the $I$=0 channel. This can be checked by looking
for an event excess in $W^+W^+ \to W^+W^+$, which contains only $I$=2.
For $WW\to t\bar t$, we assumed that the resonance was in the $I$=0
channel.  In principle, that could be checked by $t$ quark helicity
analysis, as we have explained in Section V.

We have displayed the results of this analysis in the fourth and fifth
columns of Table~\ref{tab:Peskin}. We also show in this table  results
for the annihilation reactions $q\bar q\to WZ$, $\ee\to W^+W^-$.   In
the $q\bar q$ annihilation process, the signal is a narrow resonance
peak.  We have quoted in the last column our estimate of 
the width of this peak close to 
 where it disappears into the background.  For the $\ee$
annihilation process, we have estimated the mass sensitivity from the
results of \cite{barklow}, assuming a vector meson dominance
parametrization for the new pion form factor.

\subsubsection{Effective Lagrangian Studies}

Two studies done for Snowmass used an  effective chiral Lagrangian to
parametrize the $\pi\pi$ scattering amplitudes \cite{Pelaez, Kilian}.
They then reported constraints on the parameters of the chiral
Lagrangian that could be obtained at future colliders.  Though these
analyses use a different language that the one we have used up to this
point, it is instructive to convert their results to our
parametrization where this can be done simply, in order to check the
results of the previous subsection.

The effective Lagrangian method introduces as its variable an $SU(2)$
unitary matrix $U$.  The expectation value of this matrix is nonzero if
$SU(2)$ is spontaneously broken.  The matrix $U$ is related to the pion
field by
\beq 
                   U = e^{-i\pi^a \sigma^a/2v},
\eeq{Udefin}
where $v$ is given by \leqn{fpi}.  The most general Lagrangian function
of $U$ which is invariant under $SU(2)\times SU(2)$ chiral symmetry 
is \cite{GL}
\beqa
\L &=& {v^2\over 4}\tr[\del_\mu U^\dagger \del^\mu U]  + 
      L_1 \left(\tr[\del_\mu U^\dagger \del^\mu U]\right)^2\CR & &
   \hskip 0.4in  + 
   L_2 \left(\tr[\del_\mu U^\dagger \del_\nu  U]\right)^2 + \cdots \ , 
\eeqa{genLL}
where the omitted terms contain at least six derivatives. The $\pi\pi$
scattering amplitudes derived from \leqn{genLL} do not respect
unitarity, but they are properly crossing symmetric.  The terms
proportional to $L_1$ and $L_2$ give the most general terms of order
$s^2$ which can be added to \leqn{aasym} in a chirally invariant
theory.  In \cite{Kilian}, the parameters $L_1$ and $L_2$ are called
$\alpha_5$ and $\alpha_4$.

By comparing this parametrization to \leqn{standarddelta}, one can work
out the relation between $L_1$ and $L_2$ and the masses $M_0$ and $M_1$
in that formula.  When $L_2$=0, this relation takes the simple form
\beq
         M_0^2  = {v^2\over  8 L_1} \ .
\eeq{Mzerorel}
Then we can convert the results of these papers to our terminology as
follows.  Pel\'aez \cite{Pelaez} studied the reaction $qq\to qq ZZ$ at
the parton level following the scheme of cuts of the CMS collaboration
\cite{cms}.  He found that the 95\% confidence upper limit on $L_1$ in
the LET was about $3.5\times 10^{-3}$. Kilian \cite{Kilian} studied
the reactions $\ee\to \nu\bar\nu ZZ$ and $\ee\to \ee W^+W^-$ at the
parton level for a 1.6 TeV $\ee$ collider, following the scheme of cuts
suggested in \cite{Barger}. His results were presented for 200
fb$^{-1}$ and unpolarized beams; scaling to 100 fb$^{-1}$ and 80\%
polarization, we find a  95\% confidence upper limit on $L_1$ of about
$1.8\times 10^{-3}$.  Converting these values to  limits on $M_0$, we
find the two entries in the extreme right-hand column of
Table~\ref{tab:Peskin}.

\subsubsection{Results}

We were surprised at the quality of the information that the set of
experiments shown in Table~\ref{tab:Peskin} will make available.  The
experiments are sensitive to resonances in each of the three partial
waves available to new pion scattering, up to resonance masses
significantly larger than the unitarity limits shown in \leqn{ulimits}.
We have discussed many qualifications of this analysis, most
importantly, our assumption that a single resonance dominates each
partial wave.  However, under the assumptions we have made the masses
of the resonances are obtained to 10-15\%.  And it is possible to test
whether an $I$=0 resonance which appears in $WW\to t\bar t$ is the same
one that appears in $WW\to ZZ$.

\subsection{Further Questions}

There are many questions that still must be addressed, and which
require analysis beyond the level possible in a summer study.  We list
the most important open issues here.

First of all, the analysis we have done to determine the sensitivity to
resonance masses of the new strong interactions should be repeated
using detailed simulations, at least at the parton level, to produce
the correct average over $\pi\pi$ reactions of different center-of-mass
energy.

For the LHC studies, two major questions could benefit from future
work. Since the $\pi\pi$ scattering signals are seen as event excesses
above background, it is necessary that the backgrounds be estimated
confidently.  Backgrounds which depend on missing transverse momentum
and thus on cracks and gaps in the detector need experimental
calibration.  In addition, the efficiencies of jet tags and jet vetos
must be determined experimentally, and strategies are needed to do this.

In addition, we feel it is important to extend the LHC simulation
studies beyond the study of the lowest-background `gold-plated' modes. 
Once the existence of new strong interactions has been established and
we are interested in measuring parameters of the new strong sector, it
will be important to have probes of this sector that allow
higher-statistics measurements even at the cost of new uncertainties in
backgrounds.  More work is required on studies of the `silver-plated'
modes, with one leptonic and one hadronic weak boson decay, with
forward jet tagging. CMS has already shown promising results in a 1 TeV
Higgs search with this strategy.  Another means of controlling
background is to associate $WW$ scattering events with  rapidity gaps
or other measures of low central hadronic multiplicity.  Here it seems
especially important to find a way to calibrate the tag experimentally.

For the NLC studies, though we believe that the major effects which
determine the experimental yields and resolutions have been taken into
account, it would be best to repeat the analyses of boson-boson
scattering using realistic simulations of hadronization, detector
effects, and beamstrahlung.  These simulations could also be used to
explore additional observables available at the NLC, in particular, the
use of $W$ and $Z$ helicity analysis such as we discussed for
annihilation processes in Section IVC2.  Positron polarization would
also enhance the capabilities of the NLC, and this should be studied.

In addition, it is interesting to think about the desired energy of a
future lepton collider.  The case of a 4 TeV $\ee$ or $\mu^+\mu^-$
collider has been studied \cite{BBGH} and, not surprisingly, it
provides significant increases in the projected signals over those
available from LHC or the 1.5 TeV NLC. We will see on what time scale
this technology could become available. For the near term, there is a 
question whose answer is less obvious: Would we gain from extending the
reach of the 1.5 TeV $\ee$ collider to 2 TeV with an attendant loss in
luminosity?  The optimization of energy versus luminosity for
boson-boson studies with $\ee$ colliders needs further study.

\subsection{Issues for Very High Energies}

At the beginning of  Section  VIB, we stated that the detailed mapping
of the resonances in the new strong interactions was not a realistic
goal for the next generation of colliders. However, if there are new
strong interactions, we will eventually need to face this problem.  To
our knowledge, this issue has received almost no consideration, even
though it may become the major focus of experimental particle physics
two decades from now.

At an $\ee$ or $\mu^+\mu^-$ collider, the problem of mapping resonances
is relatively straightforward.  If the constituents of the new strong
interactions have $SU(2)\times U(1)$ charges, which they must to
produce electroweak symmetry breaking, the $J$=1 resonances of this
sector will appear in lepton-antilepton annihilation.  To study these
resonances, one needs the ability to produce data samples which grow as
the square of the energy; for example, samples of 1000 fb$^{-1}$ are
needed at $\sqrt{s}$ = 5 TeV.  The new resonances presumably decay
dominantly to new pions, that is, to $W$ and $Z$ bosons.  The most
direct experimental strategy would be to identify these bosons in their
hadronic decay modes by making 2-jet mass combinations. This raises an
interesting question: In  $\ee$ reactions at present energies,
annihilation processes at the full $\sqrt{s}$ can be distinguished from
two-photon processes and other peripheral reactions by cuts on total
energy and visible transverse energy. Does this strategy continue to
work at center-of-mass energies of order 5 TeV when one includes the
new sources of gauge boson production from photon-photon collisions and
direct $W$ and $Z$ radiation?  This question should be studied in
simulations.

For a $pp$ collider at 100 TeV in the center of mass, it is less
obvious what the strategy would be to explore the resonance region of
the new strong interactions. The one part of this question that has
been addressed in the literature is that of the observability of
multiple $W$ production due to new strong 
interactions \cite{CGWWW,DMor}.  A
100 TeV collider would allow boson-boson collisions at energies
corresponding to the multiple pion production region  of QCD. In a
preliminary study at Snowmass, Kilgore and Peskin asked whether the
multiple production of new pions (observed as $W$ and $Z$ bosons with
$p_T$ above 500 GeV) would be visible over standard model backgrounds. 
Because it decreases the rate too much to insist that all of the weak
bosons decay leptonically, most of the weak bosons must be identified
in their hadronic decay modes and thus will resemble high $p_T$ jets.
The dominant backgrounds come from QCD production of $W$ or $Z$ plus
multiple gluon jets.  The signature is the presence of events with 
high jet multiplicity per unit rapidity, with $dn/dy \sim 3$ for new
pion production versus $dn/dy \sim 0.2$ for the standard model
reaction.  They estimated that a data sample of order 1000 fb$^{-1}$
would be necessary to study this process.  If QCD is a guide, the study
of the truly short distance regime of the new strong interactions would
presumably require several further orders of magnitude in luminosity.

This analysis gives one example of the type of question that will need
to be addressed to define the experimental program of these very high
energy colliders.  The formulation of this experimental program is
clearly at an early stage.

\section{Anomalous $W$ couplings}

Up to this point, we have discussed probes of the new strong
interactions through scattering processes of new pions.  An alternative
way to probe these new interactions is to study their effect on the
couplings of $W$ and $Z$ bosons.  For example, the standard Yang-Mills
coupling of the $W$ boson to the photon may be modified by
\beq
\Delta \L = e \left[ \left( W^\dagger_{\mu\nu} W^\mu -
     W^{\dagger\mu} W_{\mu\nu} \right) A^\nu + 
\kappa_\gamma W^\dagger_\mu W_\nu F^{\mu\nu} + \cdots \right]
\eeq{theLWW}
The standard Yang-Mills coupling is given by $\kappa_\gamma =1$.  A
deviation of $\kappa_\gamma$ from this value signals an anomalous
magnetic moment for the $W$ boson. The complete phenomenology of the
possible modifications of this form is described in the contribution
\cite{anom} to these proceedings.  Here, we will make only a few
general remarks which connect that study with the theory of pion-pion
interactions.

\begin{figure}[hbt]
\leavevmode
\begin{center}
\resizebox{2.1in}{!}{%
\includegraphics{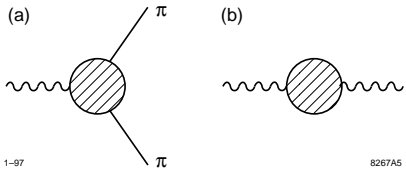}}
\end{center}
\caption{Corrections to gauge boson vertices from 
new strong interactions.}
\label{fig:WBpi}
\end{figure}

To understand how new pion interactions affect the $W$ and $Z$
couplings, the most straightforward method is to write the most general
effective Lagrangian for pions coupled in a gauge-invariant way to
$SU(2)\times U(1)$. Some time ago, Gasser and Leutwyler  \cite{GL}
addressed this problem for the conventional strong interactions,
considering arbitrary vector bosons coupling to the $SU(2)\times SU(2)$
global symmetry. Using the analogy between QCD and the new strong
interactions, and specializing their results to the standard model
gauge couplings, one finds as the leading corrections
\cite{BDV,FLS}
\beqa
    \delta\L &=& -iL_{9L}\epsilon^{abc} gW_{\mu\nu}^a \del^\mu \pi^b \del^\nu 
       \pi^c \CR & & 
-i L_{9R} \epsilon^{3bc} g'B_{\mu\nu} \del^\mu \pi^b \del^\nu \pi^c \CR & &
     + L_{10} B_{\mu\nu} W^{\mu\nu 3}, 
\eeqa{GLWcorrect}
where $a,b,c = 1,2,3$ and $W_{\mu\nu}^a$, $B_{\mu\nu}$ are the field
strength tensors of $SU(2)$ and $U(1)$ gauge fields, and the $L_i$ are
new phenomenological parameters. As shown in Figure~\ref{fig:WBpi}, the
first two terms of \leqn{GLWcorrect} can be thought of as resulting
from new strong interactions at the gauge boson vertex. The third term
can be thought of as the contribution to the vacuum polarization
resulting from virtual states of the new sector.

To understand the importance of the new couplings introduced in
\leqn{GLWcorrect}, we should first estimate the $L_i$ and then relate
these parameters to quantities such as $\kappa_\gamma$ that express the
corrections to the $WW\gamma$ and $WWZ$ vertices.  The simplest way to
estimate the $L_i$ is to saturate the strong interaction amplitudes
shown in Figure \ref{fig:WBpi} with low-lying resonances.  Since this
is just an estimate, we will assume for simplicity that the new strong
interactions conserve parity; this sets $L_{9L} = L_{9R}$.  Then, if
$M_1$ is the mass of the lowest $I$=1 vector resonance, 
\beqa
            L_9 &=& \half {f_\rho^2\over M_1^2} \CR
           L_{10} &=& - \half {f_\rho^2\over M_1^2}
 \left( 1 - {f_\rho^2M_1^2\over f_A^2 M_{A1}^2} \right) 
\eeqa{Lnineten}
where $M_{A1}$ is the mass of the lowest $I$=1 axial vector resonance.
In a more sophisticated estimate, these expressions would be replaced by 
dispersion relations over the spectrum of vector and axial vector 
mesons \cite{GL,PandT}.

The parameter $L_{10}$ is actually already known, since it is related to 
the $S$ parameter of precision electroweak physics, by
\beq
           L_{10} = - {1\over 16 \pi} S \ .
\eeq{StoLval}
The value of $S$ given in \leqn{Svalue} leads to $L_{10} = + (5\pm
3)\times 10^{-3}$.  This would seem to imply an unconventional ordering
or degeneracy of the vector and axial vector mesons of the new sector
relative to the quark model expectation.

Given values for the $L_i$, the parameters $\kappa_\gamma$ and
$\kappa_Z$ expressing the corrections to the $WW\gamma$ and $WWZ$
vertices are given by
\beqa
    \kappa_\gamma &=&  1 - 4\pi\alpha_w \cdot \half \bigl(L_{9L} + 
             L_{9R} + 2 L_{10} \bigr) \CR
    \kappa_Z &=&  1 + 4\pi\alpha_w \bigl(L_{9L}  
         -{\sstw\over \cstw}    L_{9R} \CR
              & & \hskip 1.0in + {2\over \cstw} L_{10} \bigr) \ .
\eeqa{kappavals}
In addition, the overall strength of the $WWZ$ vertex is shifted by a 
factor
\beqa
       g_{1Z} &=& + 4\pi\alpha_w \cdot \bigg({1\over 2\cstw}L_{9L}  \CR
         & & \hskip 0.2in
         +{\sstw\over \cstw(\cstw-\sstw)} L_{10} \bigg) \ .
\eeqa{goneZval}

In the previous section, we argued that the pion-pion scattering
effects can show up in the new pion form factor.  In cases where this
form factor does not contain a resonance and we observe just the first
deviation from its low-energy value, this deviation is a special case
of the couplings described in this section.  Specifically,
\beq
          \hbox{\rm Re}\ F_\pi(q^2) = 1 + {s\over f_\pi^2} \left (L_{9L}
                 + L_{9R} \right) + {\cal O}(s^2) \ .
\eeq{FpifromL}
(The imaginary part of $F_\pi$ is generated by loop corrections to the
pion effective Lagrangian.)

If we estimate $L_{9L}$, $L_{9R}$ by evaluating \leqn{Lnineten} with
the parameter values of QCD, and we ignore $L_{10}$ in accord with the
electroweak data, we find
\beq
    \kappa_\gamma -1  \sim -3\times 10^{-3}  \qquad  
  \kappa_Z -1  \sim -2\times 10^{-3} \ .
\eeq{kgZvals}
Experiments which claim to probe new strong interactions through the
anomalous $W$ couplings should be designed to achieve a level of accuracy
which would be sensitive to such small corrections.

\begin{figure*}[t]
\leavevmode
\begin{center}
\resizebox{!}{2.5in}{%
\includegraphics{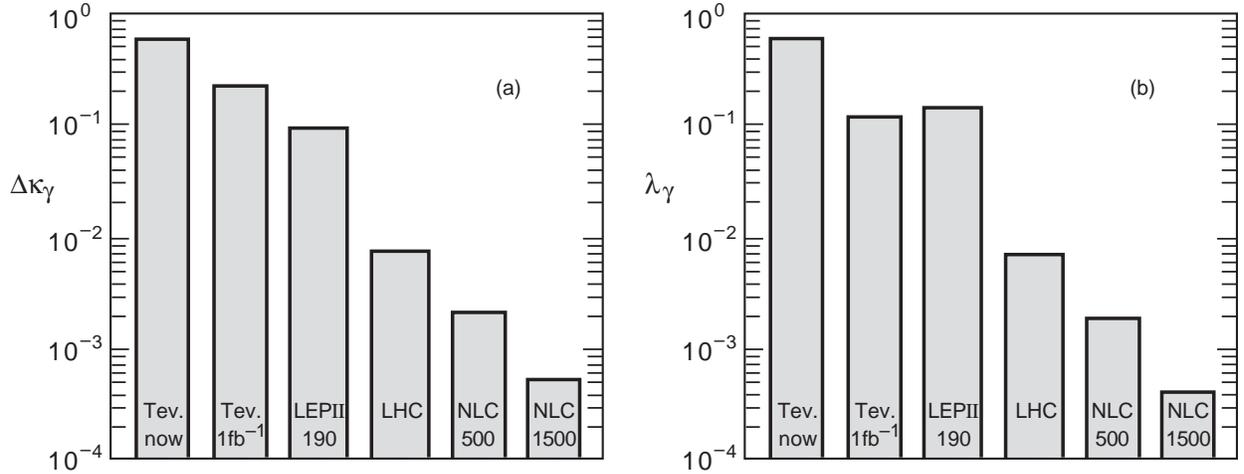}}
\end{center}
\caption{Comparison of 95\% confidence limits on the anomalous $WW\gamma$
couplings (a) $\kappa_\gamma$ 
and (b) $\lambda_\gamma$ expected from various future colliders, 
 from \protect\cite{cltp}.}
\label{fig:ac}
\end{figure*}
 
Many studies have been done to explore the sensitivity of current,
future, and proposed machines to anomalous gauge boson couplings.  The
recent DPF study of physics beyond the standard model presented the
summary of the sensitivity of future colliders to the anomalous
couplings shown in Figure~\ref{fig:ac} \cite{cltp}.  These limits are
based on a specific two-parameter representation of the anomalous
couplings \cite{HISZ}, but they illustrate well the constraints that
will be available. The studies presented at Snowmass \cite{anom} have
refined the limits  presented in the figure but do not significantly
change the projections. From a comparison of \leqn{kappavals} and
Figure~\ref{fig:ac}, we see that anomalous couplings are not
particularly sensitive probes of new strong dynamics.  The only machine
that will have sufficient sensitivity to probe the new strong
interactions through the anomalous $W$ coupling is the $e^+e^-$ linear
collider, with a study of the helicity amplitudes in  $e^+e^- \to
W^+W^-$. However, this is the same analysis and the same data that
yield a measurement of the phase shift in the $I$=1 weak isospin
channel and gives direct evidence for strong scattering.  Our
conclusion is that while experiments searching for anomalous gauge
boson couplings are extremely important as tests of the standard model,
they are not sensitive probes of strong-coupling electroweak symmetry
breaking.

\section{Fermion Masses and Model-Dependent Signatures}

Up to this point, we have been concerned with the general phenomenology
of strong interactions associated with electroweak gauge symmetry
breaking. We have discussed experimental probes of these new strong
interactions which apply irrespective of their detailed properties and
which give the most important qualitative information about their
structure.  Any successful theory of electroweak symmetry breaking,
however, must {\em also} explain the origin of the masses of the
ordinary fermions. The mechanism of fermion mass generation typically
leads to the prediction of new particles and interactions with the
property of coupling most strongly to heavy flavors. In this section,
we will discuss model-dependent experimental signatures of these new
interactions. The effects we will review are associated with specific
features of models, and so we cannot be sure that experiments which may
exclude these effects give general constraints on electroweak symmetry
breaking and fermion mass generation.  However, we will see that
explicit models of fermion mass generation often can be probed in ways
quite different from the high-energy pion-pion reactions that we have
discussed up to this point.

This section is built around a description of two features which may be
incorporated into  models of strong interaction electroweak symmetry
breaking with fermion mass generation: walking technicolor and top
quark condensation.  We will briefly outline the motivation of these
ideas and show how, in each case, the basic theoretical assumptions
give rise to new observable phenomena.  To introduce this discussion,
we review the minimal model of technicolor, the simplest scenario for
electroweak gauge symmetry breaking and fermion mass generation, which
is ultimately unsuccessful.  We will find the more realistic models by
repairing the difficulties of this one.

The concentration of this section on one particular route to model-building
may reflect the primitive state of our knowledge of  nonperturbative
phenomena in 
gauge theories, which provide the theoretical raw materials for model
construction.  Thus, it is important to look at the effects discussed in 
this section as examples of what might be found rather than predictions of
the general notion of strong-coupling electroweak symmetry breaking.
Eventually, experiments will tell us whether our current theoretical ideas
are good enough, or whether they must be expanded.

\subsection{Technicolor and Extended Technicolor}

The ideal model of new strong interactions responsible for $SU(2)\times
U(1)$ is technicolor \cite{Wein,Suss}. In this model, one postulates a
new set of interactions with exactly the physics of QCD, but with QCD
scale $\Lambda$ set several thousand times larger in accord with
\leqn{fpi}.  If the model contains two flavors of massless
`techniquarks' $T$, it has $SU(2)\times SU(2)$ global symmetry.  This
symmetry is spontaneously broken in the correct pattern through quark
pair condensation and dynamical mass generation, exactly as happens in
the familiar strong interactions.

\begin{figure}[hbt]
\leavevmode
\begin{center}
\resizebox{1.5in}{!}{%
\includegraphics{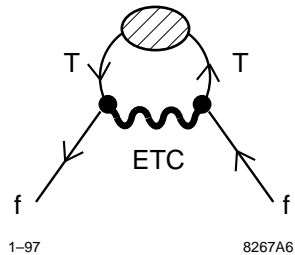}}
\end{center}
\caption{Generation of quark and lepton masses through extended technicolor
 interactions.}
\label{fig:qlmass}
\end{figure}

Masses for quarks and leptons must be induced by additional
interactions that couple these fermions to the techniquarks.  A
plausible hypothesis is that these couplings result from further new
gauge interactions, called `extended technicolor' (ETC) \cite{EL,DS}. 
The theory then produces quark and lepton masses proportional to the
dynamical mass of the techniquarks, through the diagram shown in
Figure~\ref{fig:qlmass}.

Though this scenario is simple and compelling, it cannot be correct. 
It has two difficulties associated with the ETC mechanism, and one
associated with precision electroweak results.  First, since the ETC
interactions must distinguish between flavors, the ETC gauge group
typically contains vector bosons that couple to the first two
generations  and mediate flavor changing neutral current processes such
as $K^0$--$\bar K^0$ mixing, at an unacceptable level \cite{EL,ELP,DE}.
Though these bosons are very heavy, typically in the mass range of 50
TeV, the constraints on flavor-changing neutral currents are very
stringent.  Second, the ETC mechanism cannot give a large mass to the
top quark without fine-tuning \cite{ABCH,strongETC}.  For top quark
masses above about 100 GeV, the mass of the required ETC boson comes
down to about 1 TeV, where this particle would play a role in the
technicolor dynamics and, in particular, spoil custodial $SU(2)$
symmetry.  Finally, because this theory is a scaled-up version of QCD,
the value of $S$ conflicts with precision electroweak results as we
have discussed above.

Despite these problems, many authors have viewed technicolor as a
reasonable starting point for a theory of electroweak symmetry breaking
induced by new strong interactions.  The question is, what should be
changed so that this mechanism avoids its specific problems?  We now
consider two possible answers.

\subsection{Walking technicolor}

Much of the analysis above is based on our understanding of QCD. As one
scales higher in energy, QCD goes rather suddenly to a weakly coupled
theory due to asymptotic freedom.  The first proposal, ``walking
technicolor'' \cite{Holdom,Htwo,YBM}, assumes that the short-distance
behavior of the technicolor interactions is instead dominated by an
ultraviolet fixed point. In this case, or even if the approach to
asymptotic freedom is very slow \cite{AKW}, the dynamics of the strong
interactions could be quite different.

Recently, it has been found that the supersymmetric generalization of
QCD, for a large number of quark flavors, has a nontrivial fixed point
with a manifold of vacuum states, including a point at which chiral
symmetry is unbroken \cite{Seiberg}.  In walking technicolor, we hope
that  either an approximate \cite{AppTerWij} or exact fixed point
governs the asymptotic ultraviolet behavior of ordinary,
nonsupersymmetric, QCD with a large number of flavors, or with matter
in some nonstandard gauge representation. However, to obtain
electroweak symmetry breaking, we must also assume that chiral symmetry
is broken according to the standard pattern.

The consequences of the new short-distance behavior may include
nonperturbative short-distance enhancements of amplitudes which involve
expectation values of techniquark bilinears.  In particular, the
process shown in Figure~\ref{fig:qlmass} is enhanced, pushing up the
general scale of ETC bosons and thus suppressing the dangerous
flavor-changing neutral current amplitudes involving the first two
generations.

This proposal has important experimental consequences. If the
short-distance behavior of technicolor is due to the presence of many
technifermions, this predicts the existence of additional pseudoscalar
mesons which are composites of the techniquarks.  In QCD, when we
consider a theory with a light strange quark, we add to the pion
isotriplet the kaons and the eta; similarly, a generalization of QCD
with $n$ flavors leads to $(n^2-1)$ meson species, including the three
pions.  The new particles are known as `technipions' or
`pseudo-Goldstone bosons' $P$.

In models where techniquarks carry QCD color, we can form technicolor
singlet mesons which carry QCD color, for example, by combining two
colored techniquarks into a technicolor-singlet, color-octet
combination.  However, in all models, some techniquarks must transform
under weak interaction $SU(2)$ and custodial $SU(2)$, and therefore the
additional mesons can form isospin doublets, triplets, and possibly
also higher isospin representations.  For example, a model which
contains three doublets of technifermions which transform as color
triplets $(U,D)$ predicts 35 meson states: 3 pions and an isotriplet of
color octets $P^{a+}, P^{a0}, P^{a-}$, and a color octet isosinglet
$P^{a0\prime}$.  The ETC-interactions associated with the top quark
mass would allow these particles to decay preferentially to
third-generation fermions pairs such as $t\bar t$, $t\bar b$, and
$t\tau$.

The technipion masses come from two sources.  First, there is a
contribution from standard model gauge interactions.  This is the
technicolor analogue of the $\pi^+$--$\pi^0$ mass difference; its value
is about 300 GeV for colored states \cite{us}.  Second, there is a
contribution from ETC interactions of the techniquarks inside the
mesons.  This contribution is small in conventional technicolor, but in
walking technicolor it receives a double enhancement and is expected to
be the dominant contribution \cite{Holdom}.  Unfortunately, this
contribution cannot be accurately calculated even in the simplest
realistic models.

If the technicolor model contains techniquarks in higher technicolor
representations, it is possible that the techniquarks might have
significantly different values of their dynamical masses.  That is,
these models may contain multiple, and significantly different, scales
of chiral symmetry breaking~\cite{ELmult,LR,ELlater}. Models with
``multiscale technicolor'' have the general experimental signatures of
walking technicolor models, necessarily including large multiplets of
technipions.  In addition these models contain relatively light $I$=1
resonances, the $\rho$ resonances in the sectors of the theory
associated with the lowest chiral-symmetry breaking scale. These
resonances, which may be as light as 500 GeV \cite{LR}, can be produced
in fermion-antifermion annihilation as discussed above and provide
enhancements in the cross sections for $W^+W^-$ and $P^+P^-$
production at relatively low energies.

Finally, as emphasized by Lane \cite{lanetasi}, the  estimate of $S$
which we have given in Section IIA depends crucially on the assumption
that the new strong dynamics is QCD-like. We have explained in Section
VII how a different hadron spectrum can lead to a vanishing or negative
value.  Unfortunately, it is not known whether a walking technicolor
model can in fact have the spectrum required.

\begin{figure}[hbt]
\leavevmode
\begin{center}
\resizebox{2.0in}{!}{%
\includegraphics{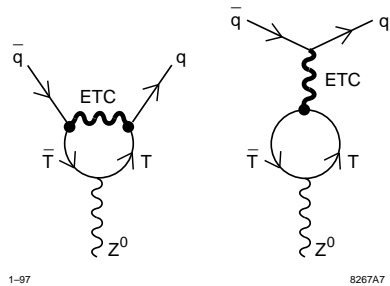}}
\end{center}
\caption{Modification of the $b\bar b Z$ and $t \bar t Z$ vertices by
    ETC interactions.}
\label{fig:Zvert}
\end{figure}

While walking technicolor mitigates the flavor-changing neutral-current
problems which arise in generating the masses of the first two
generations of quarks, it does not address the problems associated with
a heavy top quark.  In the absence of fine-tuning \cite{whofortop}, the
ETC boson associated with the top quark would still have a mass of
order 1 TeV \cite{CGST}, and in fact, the new asymptotic behavior tends
to enhance isospin-violating effects \cite{whofortop2}. An additional
potentially dangerous consequence of a light ETC gauge-boson is the
presence of anomalous couplings of the third-generation quarks. The ETC
gauge group includes gauge bosons which modify the $t$ and $b$ vertices
with the $Z^0$ through the processes shown in Figure~\ref{fig:Zvert}. 
These diagrams have been estimated \cite{CSS} and turn out to be of
order
\beq
    \delta g_Z \sim  g_Z \cdot  {m_t\over 4 \pi f_\pi}\ ,
\eeq{Zvertshift}
giving corrections of order 7\%. Note that this contribution involves
only one power of $(m_t/f_\pi)$, whereas most vertex corrections due to
top quark mass generation are suppressed by two powers of this ratio.
In explicit models \cite{KH,Guo,CST},  the contributions from standard
and diagonal ETC bosons shown in the figure are of opposite sign for
$b$ and of the same sign for $t$, so one expects a few-percent and
rather model-dependent effect on $\Gamma(Z^0\to b\bar b)$ but effects
of order 10\% for the top quark couplings  \cite{MuraMM,Mahant}.

The final  consequence of this class of models is the prediction of
heavy particles which are bound states of techniquarks with the light
ETC boson associated with top quark mass generation \cite{AW,AppelE}. 
These states have the quantum numbers and decay schemes of excited
third-generation fermions, for example,
\beq
  t^* \to tZ,\ bW\ ,  \qquad \tau^* \to \tau Z, \ \nu_\tau W \ .
\eeq{tstar}

\subsection{Topcolor-Assisted Technicolor}

To  address the problems  discussed at the end of the previous section,
one might introduce a new interaction specifically associated with the
top quark, in order to generate its large mass by top quark pair
condensation.  Such new interactions, introduced originally by Hill
\cite{hill, Hillttc}, are called `topcolor'.  Though it is a very
attractive idea that top quark pair condensation is the mechanism of
$SU(2)\times U(1)$ symmetry breaking, this is unlikely because the
asymmetry between the top and bottom masses would require a large
violation of custodial $SU(2)$.  To avoid this problem without
fine-tuning, the top condensation must be isolated from the technicolor
interactions which are the major source of the $W$ and $Z$ masses. 
This puts the top condensation in an additional new sector, with its
own characteristic phenomenology.

The strength of the top condensate is bounded above by a constraint
associated with the permitted violation of the relation \leqn{rhorel}
and is bounded below by limits on the strength of the top quark
coupling to new interactions \cite{Hillttc}.  For simplicity, the
interactions which induce top quark condensation are often modeled by
4-fermion contact interactions induced at a scale $M$ of order one TeV.
It may be more realistic to assume that the top quark couples to a new
sector of (broken) gauge interactions.  The strength of these new
interactions must be adjusted to be quite near the threshold for chiral
symmetry breaking; otherwise, the top quark mass jumps up to the scale
of the new interactions (of order one TeV).  Direct experimental
searches for these gauge bosons as resonances in the $t \bar t$ mass
distribution push their masses above 600 GeV. Under these constraints,
one finds a small contribution to the $W$ mass from top quark pair
condensation; this must be supplemented by a larger contribution due to
technicolor.  Thus, \leqn{fpi} is generalized to
\beq
            f_t^2 + f_\pi^2  = (246 \ \hbox{\rm GeV})^2 \ ,
\eeq{fpit}
where $f_\pi$ is the pion decay constant of technicolor and $f_t$ is a
new decay constant associated with the top condensate.  Typical values
are $f_t = 75$ GeV, $f_\pi = 235$ GeV, so that
\beq
    r_t = (f_t/f_\pi)^2 \approx  {1\over 10} \ .
\eeq{rtval}

In addition to the direct manifestations of new top interactions, there
is an interesting indirect consequence of this new physics.  To the
extent that the top condensation is decoupled from the
symmetry-breaking of the technicolor fields, the top condensate would
lead to its own separate set of Goldstone bosons, the `top-pions'
$\pi_t^+$, $\pi_t^0$, $\pi_t^-$.  These receive masses which are small
compared to the topcolor scale as the result of the coupling of the top
and bottom quarks to technicolor through ETC interactions.  However,
since ETC now plays only a small role in giving the top quark a mass,
these effects are suppressed.  If the ETC contribution to the top
quark mass is of order 5--10 GeV, the top-pions are expected to have a
mass between 150 and 400 GeV \cite{Hillttc}.

Furthermore, if the top condensate is induced by short-ranged
strong-interactions adjusted to be near the critical point for chiral
symmetry breaking, the theory predicts a $t\bar t$ composite scalar
with a mass close to the $t\bar t$ threshold \cite{CCL}.  The effective
size of this composite system, however, is $M^{-1}$, which means that
for practical purposes this scalar should be treated as an elementary
scalar field which we might call the `top-Higgs' $h^0_t$.  Thus,
topcolor leads naturally to the appearance of an additional sector of
apparently elementary Higgs bosons.

From this general outline of topcolor models, we can deduce three
experimental consequences.  First, topcolor gives a potential for
resonances in $t\bar t$ and $b\bar b$ production at high energy.  If
the topcolor interactions are a new gauge sector, composed of
`top-gluons' with an enhanced coupling to the third generation, these
should produce resonances in the reaction $q\bar q \to t\bar t$.  In
addition, realistic topcolor models contain a weakly coupled $U(1)$
gauge boson $Z_t^0$ which distinguishes the $t$ and $b$.  In order to
cancel anomalies, this gauge boson will also couple to leptons.  Thus,
it is likely to lead to resonances or interference effects in all
channels of $\ee\to f \bar f$, just as for any other type of heavy
neutral vector boson.

Second, as we have discussed above, topcolor predicts a top-Higgs
sector, a set of four bosons $\pi_t^i$, $h^0_t$ which behave like a
second scalar doublet in a multi-Higgs model.  The masses of these
particles are expected to be in the range 150-400 GeV.  Some of their
properties are those expected for conventional multi-Higgs models, but
some are different in an interesting way.  The pair production processes
\beq
      \ee \to \pi^+_t \pi^-_t \ ,  \ee \to \pi^0_t h^0_t
\eeq{pairoftoppi}
have the standard cross sections for an elementary scalar doublet.  The
process
\beq
     \ee \to Z^0 h^0_t \ ,
\eeq{Zthh}
however, occurs at a rate suppressed from the standard model rate for
$\ee \to Z^0 h^0$ by a factor $r_t$ in \leqn{rtval}.  On the other
hand, the single production processes
\beq
        \gamma \gamma,\ gg  \to h^0_t,
\eeq{singlepit}
are enhanced parametrically 
relative to the rate for a minimal Higgs boson by a factor
$1/r_t$.  Further aspects of the physics of these particles are discussed
in \cite{BurdTC}.

The third consequence of topcolor is a suppression of the process
$WW\to t\bar t$ discussed in Section V.  In the $I$=0 amplitude
\leqn{ttspinzero}, the mass $M_0$ of the scalar resonance should be
replaced by the mass of the $h^0_t$, giving a much lower resonance
position followed by a rapid fall-off.  In the $I$=1 amplitude
\leqn{ttspinone}, the parameter $\eta$ should be small, of order
$(m_b/m_t)$, that is, of the order of the fraction of the top quark
mass due to ETC and not topcolor interactions.  Topcolor provides a
concrete example of the claim we made in Section V, that the cross
section for $WW\to t\bar t$ tests directly whether the top quark
derives its mass from the new strong interactions that produce the $W$
mass, or from some other source.

\section{Experimental Searches for Exotic Particles}
 
In this section, we summarize the experimental reach of various
machines to exotica that are associated with specific models of strong
interaction electroweak symmetry breaking.  The number of models is
large and explicit studies have not been done for all of them. We
attempt here to give general guidelines for the experimental reaches
rather than report exhaustively for each conceivable model.  We
consider the search reach of the Tevatron in Run I (0.1fb$^{-1}$), Run II
(2fb$^{-1}$), and TeV33 (30fb$^{-1}$), the search reach of an
electron-positron linear collider (NLC) at 0.5TeV (50 fb$^{-1}$),
1.0TeV  (100fb$^{-1}$), and 1.5TeV  (100fb$^{-1}$), and the search
reach of the LHC (100fb$^{-1}$).

\begin{table*}[t]
\begin{center}
\caption{Mass reach in  model-dependent searches for exotica 
discussed in Section IX.  All mass limits are in TeV. The dashes
represent cases for which we could not find results in the literature.}
\label{tab:exotic}
\begin{tabular}{|c|ccc|ccc|c|}
\hline
\hline
  & \multicolumn{3}{c|}{NLC}   & \multicolumn{3}{c|}{Tevatron} &  LHC  \\
  & 0.5 & 1.0 & 1.5  & Run I & Run II &  TeV33   &     \\
\hline
Technipion  $P^{\pm}$  &  0.25&0.5&0.75  & --&0.1&--  & 0.3 - 0.4  \\
Technipion  $P^{0\prime}_8$    & -- & 0.7 & --  & 0.4-0.5 & -- & -- & -- \\
\hline
Technirho (singlet)  & 1.5& 3.0 & 4.0  &  --&0.2 & 0.4  &  1.6  \\
Technirho (octet)    & --& -- & --  & 0.5&--& 0.9  &--   \\
\hline
Topcolor $B$  & -- & 0.8? & --  & --&1.1 & 1.4  &  found \\
Topcolor $Z'$  &  \multicolumn{3}{c|}{found}  &  -- &0.9 & 1.1  &  found \\
\hline
Top Higgs &  0.18 &  0.55 & -- & --  & -- & -- & 0.7 \\
\hline\hline
\end{tabular}
\end{center}
\end{table*}

\subsection{Technipions}
The technipions in a given model can be electrically charged or
electrically neutral, and they can exist as color singlets, triplets,
or octets. At lepton colliders, the electrically charged technipions
are pair produced and can be detected  almost up to the kinematic limit
of the machine ($\sqrt{s}/2$).  The current experimental limits are
from the OPAL experiment at LEP1: $m(P^\pm) > 35$ GeV and $m(P_8^\pm) >
45$ GeV \cite{opal}.  Searching for the neutral technipions at a lepton
collider is a bit more subtle.  There are no tree level gauge couplings
so the neutral technipions are produced via a triangle diagram in
association with a monoenergetic gluon, photon or $Z$.  The production
cross sections can be sizeable for neutral technipions of mass a
significant fraction of $\sqrt{s}$.  Swartz has described the search
technique in a paper in these proceedings \cite{swartz}. A 1 TeV linear
collider has a mass reach of approximately 700 GeV for the neutral
color octet
technipion states.
 
The Tevatron is currently sensitive to the colored technipion states
(charged and neutral) only to a mass of 225 GeV \cite{cheung}. 
For a neutral, color octet technipion arising in multiscale
models, the $t\bar{t}$ cross section is a very sensitive probe, and so the
Tevatron has a mass reach for the $P_8^{0\prime}(\eta_T)$ of 400-500
GeV \cite{sekhar}. The color singlet technipions are more difficult to discover
at a hadron collider than the color octet states since the latter can be
strongly produced.  In Run II, the reach for the color singlet technipion
states will be 115 GeV.  At LHC, older studies which do not take advantage
of $b$ tagging indicate that the color octet technipions are observable 
through direct production up to a mass of about 325 GeV and that the LHC will
be able to see the charged color singlet technipions to a mass reach of 
about 400 GeV.  The same studies indicate that the neutral singlet 
technipions will be accessible to 100--150 GeV at LHC \cite{sekhar}.

\subsection{Technirho $\rho_T$}

We have discussed in detail the mass reach of the LHC and of a 1.5 TeV
linear collider for the color singlet 
$\rho_T$.  However, lower energy machines are
also sensitive to its discovery.  The linear collider operating at
$\sqrt{s}=500$ GeV can discover a $\rho_T$ up to at least 1.5 TeV via
the same helicity analysis of $e^+e^- \rightarrow W^+W^-$ that was
described earlier (and, again, with the assumption that a single
resonance dominates the form factor) \cite{NLCrep}.  In fact, by the 
same technique, LEP 2 can discover a $\rho_T$ up to 350 GeV \cite{barklow}.
 The Tevatron will
have a discovery reach of approximately 200 GeV in run II and 400 GeV
at TeV33.  As was the case 
with technipions, the color octet technirho states are more easily
accessible than the singlets
at a hadron collider.  The Tevatron currently excludes
a color octet technirho up to 500 GeV and 
 TeV33 will be sensitive to color octet $\rho_T$'s up to
a mass of about 900 GeV \cite{cheung}.
 
\subsection{Topcolor $B$ and $Z'$}
In topcolor assisted technicolor, there is a top gluon ($B$) that
couples mainly to top and bottom quarks and a massive color singlet
gauge boson ($Z'$). The mass reach for topgluons is 1.0-1.1 TeV for the
Tevatron in Run II and is 1.3-1.4 TeV for TeV33 \cite{cheung}.  The
sensitivity to the $Z'$ depends on the couplings of the $Z'$ to the
light quarks.  In the optimistic case where the coupling is comparable
to the couplings of the quarks to the $Z$, TeV33 will be sensitive to
the $Z'$ up to a mass of 1.1 TeV. The LHC should be  sensitive over the
entire range of expected masses for both the top gluon and topcolor
$Z'$ and will find them if they are there \cite{cheung,harris}.
 
There are no published analyses on the reach of NLC to discover
topgluons; the search is difficult because the topgluons couple
directly only to quarks. Some preliminary work \cite{Sather} indicates
that the topgluon can be observed at the 1 TeV NLC up to about 800 GeV, not
as a resonance but rather through  interference effects in $\ee\to
b\bar b b \bar b$. This point needs further study.   On the other hand,
the topcolor $Z'$ couples to leptons and thus gives effects similar to
those of other new $Z$ bosons. The NLC even at 500 GeV would be
sensitive to the topcolor $Z'$ up to 3 TeV, covering the  complete
range of expectations for its mass \cite{newZsum}.

\subsection{Light ETC Exotica}

Heavy particles which are are bound states of techniquarks with light
ETC bosons ($\tau^*$ and $t^*$) will behave much like excited quarks
and leptons.  For excited  quarks, the reach of TeV33 is 1.1 TeV and
that of  LHC is 6.3 TeV. For NLC(0.5/1.0) the reach is 0.45/0.9~TeV
\cite{NLCrep,cheung}. These limits can be used as an  estimate of the
expected reach for the $t^*$ and, in the case of the NLC, for the
$\tau^*$.
 
\subsection{Top Higgs Sector}

The $\pi_t$ will be pair-produced through Drell-Yan production, with
the dominant decays into weak boson pairs.  The $h_t$ is produced in
much the same processes as a standard model Higgs, although the width
and the branching fractions may be very different. At the NLC, a lower
limit to the discovery reach comes from considering the conventional
Higgs reaction $\ee\to Z^0 h^0$ with 10\% of the standard cross
section; this gives a reach of 180 GeV at 0.5 TeV and about 550 GeV at
1.0 TeV \cite{NLCrep}.  The latter reach is well above the expected
value of $2m_t$. (In addition, the $h_t$  should be visible as a
resonance at a $\gamma\gamma$ collider.)  We can obtain a similar
lower limit to  the discovery reach at the LHC by comparing the 
rate for the process $gg\to h_t \to Z^0 Z^0$ to that for the 
Standard Model Higgs; this gives a reach of about 700 GeV~\cite{jwells}. 
 Neither of 
these estimates makes use of the dominant decay of the top Higgs to 
$t\bar t$.  More detailed, model-dependent assessments of the top
Higgs sector discovery limits need to be made for all colliders.
 
The model dependent reach limits for the various explicit models of
discussed in Section VIII are summarized in Table \ref{tab:exotic}.

\section{Conclusion}

In this report, we have described two  approaches to the experimental
study of a strong-coupling mechanism of electroweak symmetry breaking.
In the first few sections, we took the conservative point of view that
any model of strong-coupling electroweak symmetry breaking must result
in pion-pion strong interactions that could be observed in high energy
$WW$ scattering.  We summarized the available experimental studies that
have explored the reach of future experiments to uncover and study
these new pion interactions, and we evaluated both the ability of the
experiments to discover the new strong interaction, as well as the
reach of experiments to probe the structure of the new strong
interactions.  We then turned to the study of specific models and
discussed the wide variety of signatures which characterize them.  In
general, the models predict new particles which decay to heavy flavors
and exotic interactions of the third-generation fermions.  Thus, even
if there is no single signature which is characteristic of the broad
class of models, it is clear what general experiments one must do to
probe this physics.

\begin{figure}[hbt]
\leavevmode
\begin{center}
\resizebox{!}{2.0in}{%
\includegraphics{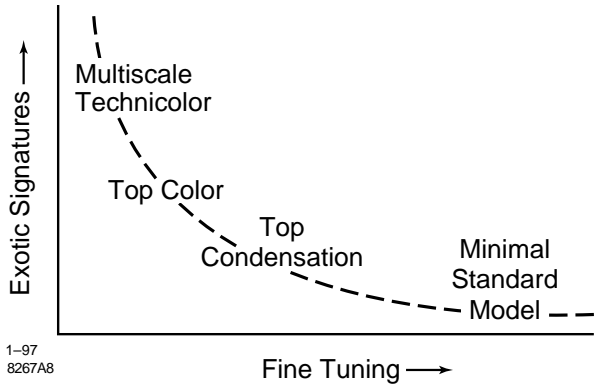}}
\end{center}
\caption{Diagram of strong-coupling model space, after Terning.}
\label{fig:ExFt}
\end{figure}

The tension between conservative, model-independent signatures on one
hand and exotic signatures on the other is illustrated in
Figure~\ref{fig:ExFt}.  The axes in this figure are, vertically, the
appearance of exotic signatures and, horizontally, the amount of
fine-tuning implicit in the model.  The scale is probably logarithmic
in both cases. Models with a large number of wheels and gears naturally
create a large number of exotic phenomena that one could try to observe
at colliders. The further one goes in removing explicit dynamical
mechanisms, the more one must rely experimentally on the most general,
model-independent signatures.  But also in this case, one must depend
more and more on fine adjustments of  the model to produce all of the
physics which results from electroweak symmetry breaking.

\begin{figure*}[hbt]
\leavevmode
\begin{center}
\resizebox{5.0in}{!}{%
\includegraphics{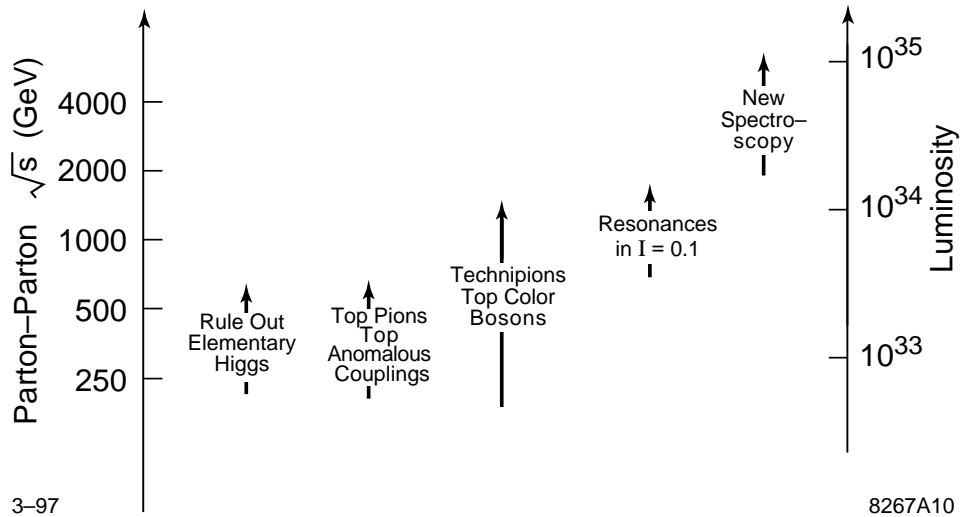}}
\end{center}
\caption{Program of experiments on strong-coupling electroweak symmetry
breaking as a function of available energy and luminosity.}
\label{fig:Lumscale}
\end{figure*}

In this report, we have listed a large number of model-dependent
signatures of new strong interactions.  The searches for these
particles can begin at present accelerators and at the upgraded
Tevatron.  The demonstration that electroweak symmetry breaking comes
from new strong interactions should also include the exclusion of a
light Higgs boson coupling to $WW$ and $ZZ$. In
Figure~\ref{fig:Lumscale}, we indicate the experiments that will be
important to the study of new strong interactions as the available
energy and luminosity is  increased.

However, the irreducible experimental question is that of the direct
observation of new strong interactions in $WW$ scattering processes. 
These experiments can be carried out at the LHC and can be confirmed
and extended by a high-energy  electron-positron collider. However, it
will require pushing those machines to the limits of their design.

It is clear that should new strong interactions be discovered, we would
then want to study their behavior at energies high compared to their
natural scale.  This study, which would be necessary to clarify the
fundamental origin of the new interactions, would require accelerator
facilities beyond the next generation---parton center-of-mass energies
in excess of 5 TeV and luminosities in excess of $10^{35}$. This study
is not a realistic goal for currently feasible accelerators, and, even
to imagine carrying out such an experimental program, it will be
essential to first know that this is the path that Nature has chosen. For
the near future, the crucial questions we must answer are simpler:  Are
there new strong interactions responsible for electroweak symmetry
breaking?  What is their general form?  Where are their most important
resonances?  How does the top quark couple to these interactions, and
why is the top quark so heavy?  These are the questions that the
working group focussed on.  Our conclusions can be summarized as
follows:
 
\begin{itemize}
\item The experiments are hard.
\item We will need the highest energy and the full design luminosity of
the proposed and upcoming colliders.  Our conclusions assume the LHC at
14 TeV and the NLC at 1.5 TeV, in each case with 100 fb$^{-1}$ data
samples.
\item The LHC can discover strong interaction electroweak symmetry
breaking.
\item The LHC can probe the effects of scalar resonances up to 1.6 TeV
and vector resonances up to 1.6 TeV.
\item The NLC at 1.5 TeV can probe the effects of resonances in the
vector channel up to 4 TeV through an accurate measurement of the $I$=1
phase shift in $e^+e^-\to W^+W^-$. 
\item The NLC at 1.5 TeV can study the coupling of top to the symmetry
breaking sector. 
\item Many, but not all, models of strong electro-weak symmetry
breaking predict low energy phenomena in the mass range 200--700 GeV
that can be searched for at the Tevatron, LHC, and NLC. 
\item A detailed understanding of strong electro-weak symmetry breaking
will probably have to wait for machines with parton energies greater
than 5 TeV and luminosities greater than $10^{35}$ cm$^{-2}$s$^{-1}$.
\end{itemize}
 
%

\end{document}